\documentclass[aps,preprint,floats,epsf,epsfig,nofootinbib,letter]{revtex4}

\usepackage{amsmath}
\usepackage{color}
\usepackage{graphicx}
\usepackage{dcolumn}
\usepackage{bm}

\def\ol{\overline}
\def\ov{\overline}
\def\be{\begin{eqnarray}}
\def\en{\end{eqnarray}}
\def\non{\nonumber}
\def\CP{{\it CP}~}
\def\vma{{_{V-A}}}

\def\la{\langle}
\def\ra{\rangle}

\def\B{{\cal B}}

\def\PAP{{P\!A_P}}
\def\PAV{{P\!A_V}}
\def\PEP{{P\!E_P}}
\def\PEV{{P\!E_V}}
\def\PE{{P\!E}}
\def\PA{{P\!A}}
\def\lsim{ {\ \lower-1.2pt\vbox{\hbox{\rlap{$<$}\lower5pt\vbox{\hbox{$\sim$}
}}}\ } }
\def\gsim{ {\ \lower-1.2pt\vbox{\hbox{\rlap{$>$}\lower5pt\vbox{\hbox{$\sim$}
}}}\ } }

\begin{document}

\font\el=cmbx10 scaled \magstep2{\obeylines\hfill October, 2019}
\vskip 1.0 cm

\title{Revisiting {\it CP} violation in $D\to P\!P$ and $V\!P$ decays}

\author{Hai-Yang Cheng}
\affiliation{Institute of Physics, Academia Sinica, Taipei, Taiwan 11529, ROC}

\author{Cheng-Wei Chiang}
\affiliation{Department of Physics, National Taiwan University, Taipei, Taiwan 10617, ROC}

\bigskip
\begin{abstract}
\bigskip
Direct \CP violation in the hadronic charm decays provides a good testing ground for the Kobayashi-Maskawa mechanism in the Standard Model.  Any significant deviations from the expectation would be indirect evidence of physics beyond the Standard Model.  In view of improved measurements from LHCb and BESIII experiments, we re-analyze the Cabibbo-favored $D \to P\!P$ and $V\!P$ decays in the topological diagram approach.  By assuming certain SU(3)-breaking effects in the tree-type amplitudes, we make predictions for both branching fractions and \CP asymmetries of the singly Cabibbo-suppressed decay modes.  While the color-allowed and -suppressed amplitudes are preferred to scale by the factor dictated by factorization in the $P\!P$ modes, no such scaling is required in the $V\!P$ modes.  The $W$-exchange amplitudes are found to change by 10\% to 50\% and depend on whether $d\overline{d}$ or $s\overline{s}$ pair directly emerges from $W$-exchange.  The predictions of branching fractions are generally improved after these SU(3) symmetry breaking effects are taken into account.  We show in detail how the tree-type, QCD-penguin, and weak penguin-annihilation diagrams contribute and modify \CP asymmetry predictions.  Future measurements of sufficiently many direct \CP asymmetries will be very useful in removing a discrete ambiguity in the strong phases as well as discriminating among different theory approaches.  In particular, we predict $a_{CP}(K^+K^-)-a_{CP}(\pi^+\pi^-) = (-1.14 \pm 0.26) \times 10^{-3}$ or $(-1.25 \pm 0.25) \times 10^{-3}$, consistent with the latest data, and $a_{CP}(K^+K^{*-})-a_{CP}(\pi^+\rho^-) = (-1.52 \pm 0.43) \times 10^{-3}$, an attractive and measurable observable in the near future.  Moreover, we observe that such \CP asymmetry differences are dominated by long-distance penguin-exchange through final-state rescattering.
\end{abstract}



\maketitle
\small

\newpage

\section{Introduction \label{sec:intro}}
Based on 0.62 fb$^{-1}$ of 2011 data, in 2012 the LHCb Collaboration has reported a result of a nonzero value for the difference between the time-integrated \CP asymmetries of the decays $D^0\to K^+K^-$ and $D^0\to\pi^+\pi^-$~\cite{LHCb2012}
\be \label{eq:LHCb2012}
\Delta A_{CP}\equiv a_{CP}(K^+K^-)-a_{CP}(\pi^+\pi^-)=-(0.82\pm0.21\pm0.11)\% \qquad {\rm (LHCb2012)}.
\en
The time-integrated asymmetry can be further decomposed into a direct \CP asymmetry $a_{CP}^{\rm dir}$
and a mixing-induced indirect \CP asymmetry $a_{CP}^{\rm ind}$
\be
a_{CP}(f)=a_{CP}^{\rm dir}(f)\left( 1+{\la t\ra\over\tau}y_{CP}\right) +{\la t\ra\over\tau} a_{CP}^{\rm ind} \ ,
\en
where  $\la t\ra$ is the average decay time in the sample, $\tau$ is the $D^0$ lifetime and $y_{CP}$ is the deviation from unity of the ratio of the effective lifetimes of $D^0$ meson decays to flavor-specific and {\it CP}-even final states. To a good approximation, $a_{CP}^{\rm ind}$ is independent of the decay mode. Hence,
\be
\Delta A_{CP}=\Delta a_{CP}^{\rm dir}\left( 1+{\overline{\la t\ra}\over\tau}y_{CP}\right) +{\Delta \la t\ra\over\tau} a_{CP}^{\rm ind} \ .
\en
Based on the LHCb averages of $y_{CP}$ and $a_{CP}^{\rm ind}$, it is known that $\Delta A_{CP}$ is primarily sensitive to direct \CP violation.

Since $\Delta a_{CP}^{\rm dir}$ in the Standard Model (SM) is naively expected  to be at most of order $1\times 10^{-3}$,  many new physics (NP) models~\cite{Isidori:2011,Zhu,Rozanov,Nir,LiuC,Giudice,Altmannshofer,Chen,Hiller:2012,DaRold,Delaunay,Chen:b,Delepine,Zupan} had been proposed to explain the measurement of large $\Delta A_{CP}$, although
it was also argued in~\cite{Kagan,PU,Brod,Feldmann,Franco,Bhattacharya:2012,Atwood,Hiller} that large \CP asymmetries in singly Cabibbo-suppressed (SCS) $D$ decays were allowed in the SM due to some nonperturbative effects or unexpected strong dynamics and the measured $\Delta a_{CP}^{\rm dir}$ could be accommodated or marginally achieved.

On the experimental side, the large $\Delta A_{CP}$ observed by LHCb in 2011 was subsequently confirmed by CDF~\cite{CDF} and by Belle~\cite{Belle}. However, the effects disappeared in the muon-tag LHCb analyses in 2013 and 2014~\cite{LHCb:2013,LHCb:2014} and were not seen in the subsequent pion-tag analysis in 2016~\cite{LHCb:2016}. Finally, in this year LHCb announced the measurements based on pion and muon tagged analyses~\cite{LHCb:2019}. Combining these with previous LHCb results in 2014 and 2016 leads to~\cite{LHCb:2019}
\be \label{eq:LHCb:2019}
\Delta A_{CP}=(-1.54\pm0.29)\times 10^{-3}, \qquad {\rm (LHCb2019)}.
\en
which yields $\Delta a_{CP}^{\rm dir}=(-1.56\pm 0.29)\times 10^{-3}$. This is the first observation of \CP violation in the charm sector!

It is most important to explore whether the first observation of \CP violation in the charm sector (\ref{eq:LHCb:2019}) is consistent with the standard model or not.
\footnote{There were a few theory papers~\cite{Xing:2019uzz,Chala:2019fdb,Li:2019hho,Grossman:2019xcj,Soni:2019xko,Nir} after the 2019 LHCb measurement.}
A common argument against the SM interpretation of Eq.~(\ref{eq:LHCb:2019}) goes as follows. Consider the tree $T$ and penguin $P$ contributions to $D^0\to K^+K^-$ and $D^0\to \pi^+\pi^-$. A simplified expression of the \CP asymmetry difference between them is given by (for a complete expression of $\Delta a_{CP}^{\rm dir}$, see Eq.~(\ref{eq:DCPVdiff}) below)
\be
\Delta a_{CP}^{\rm dir}\approx -1.3\times 10^{-3} \left( \left|{P\over T}\right|_{_{K\!K}}\sin\theta_{_{K\!K}}+ \left|{P\over T}\right|_{_{\pi\pi}}\sin\theta_{_{\pi\pi}} \right),
\en
where $\theta{_{K\!K}}$ is the strong phase of $(P/T)_{_{K\!K}}$ and likewise for $\theta{_{\pi\pi}}$. Since $|P/T|$ is na{\"i}vely expected to be of order $(\alpha_s(\mu_c)/\pi)\sim {\cal O}(0.1)$, it appears that $\Delta a_{CP}^{\rm dir}$ is most likely of order $10^{-4}$ even if the strong phases are allowed to be close to $90^\circ$. Indeed, using the results of $|P/T|$ obtained from light-cone sum rules, the authors of~\cite{Khodjamirian:2017} claimed an upper bound in the SM, $|\Delta A_{CP}^{\rm SM}|\leq (2.0\pm0.3)\times 10^{-4}$. The notion that this would imply new physics was reinforced by a recent similar analysis~\cite{Chala:2019fdb}.

In 2012, we have studied direct \CP violation in charmed meson decays based on the topological diagram approach for tree amplitudes and QCD factorization for penguin amplitudes~\cite{Cheng:2012b,Cheng:2012a}. We have pointed out the importance of a resonantlike final-state rescattering which has the same topology as the QCD-penguin exchange toplogical graph. Hence, penguin annihilation receives sizable long-distance contributions from final-state interactions. We have shown that $\Delta a_{CP}^{\rm dir}$ arises mainly from long-distance weak penguin annihilation.  Moreover, we predicted that $\Delta a_{CP}^{\rm dir}$ is about $(-0.139\pm0.004)\%$ and $(-0.151\pm0.004)\%$ for the two solutions of $W$-exchange amplitudes~\cite{Cheng:2012b}. Those were the main predictions among others made in 2012. Since the world average during that time was $\Delta a_{CP}^{\rm dir}=(-0.645\pm0.180)\%$~\cite{HFLAV}, we concluded that if this \CP asymmetry difference continues to be large with more statistics in the future, it will be clear evidence of physics beyond the standard model in the charm sector. Nowadays, we know that the LHCb new measurement almost coincides with our second solution.
This implies that one does not need New Physics at all to understand the first observation of $\Delta a_{CP}^{\rm dir}$ by LHCb! \footnote{A similar result of $\Delta a_{CP}^{\rm dir}$ based on a variant of the diagrammatic approach was obtained in~\cite{Li:2012}.}

The purpose of this work is twofold. First, we would like to improve the analysis of \CP asymmetries in $D\to P\!P$ decays. For example, it is well known that the penguin-exchange amplitude $\PE$ and the penguin-annihilation one $\PA$ evaluated in the approach of QCD factorization is subject to the end-point divergence. We need to address this issue. Also in our previous study of the long-distance contribution to $\PE$, we did not consider the uncertainties connected with final-state rescattering~\cite{Cheng:2012b}. This will be improved in this work. Secondly, although we have studied \CP asymmetries in $D\to V\!P$ decays before in~\cite{Cheng:2012a}, we focused only to the neutral charmed meson ones. Owing to the lack of information on $W$-annihilation amplitudes, no prediction was attempted for $D^+\to V\!P$ and $D_s^+\to V\!P$ decays. Thanks to the BaBar's measurement of $D_s^+\to\pi^+\rho^0$~\cite{BaBarrhopi}, the amplitudes $A_{V,P}$ can be extracted for the first time in~\cite{Cheng:2016}. Consequently, in this work we are able to complete the analysis of \CP violation in the $V\!P$ sector.

The layout of the present paper is as follows.  After a brief review of the diagrammatic approach, we study various mechanisms responsible for the large SU(3) violation in the branching fraction ratio of $D^0\to K^+K^-$ to $D^0\to \pi^+\pi^-$ and fix the SU(3) breaking effects in weak annihilation amplitudes in Section~\ref{sec:flavordiagram}. Penguin amplitudes are studied in the framework of QCD factorization as illustrated in Section~\ref{sec:penguin}. We then discuss direct \CP violation in SCS $D\to P\!P$ decays in Section~\ref{sec:DCPV} and compare our results with other works in the literature. Section \ref{sec:vp} is devoted to $D\to V\!P$ decays and their direct \CP asymmetries.
Finally, Section~\ref{sec:summary} comes to our conclusions.

\section{$D \to P\!P$ Decays \label{sec:flavordiagram}}

It is known that a reliable theoretical description of the underlying mechanism for exclusive hadronic $D$ decays based on QCD is still not yet available as the mass of the charm quark, being about $1.3$~GeV, is not heavy enough to allow for a sensible heavy quark expansion.
It has been established sometime ago that a more suitable framework for the analysis of hadronic charmed meson decays is the so-called topological diagram approach~\cite{Chau,CC86,CC87}.
In this diagrammatic scenario, the topological diagrams can be classified into three distinct groups (see Fig.~1 of~\cite{Cheng:2012a}). The first two of them (see~\cite{ChengOh} for details) are:

1. Tree and penguin amplitudes: color-allowed tree amplitude $T$; color-suppressed tree amplitude $C$; QCD-penguin amplitude $P$; singlet QCD-penguin amplitude $S$ involving flavor SU(3)-singlet mesons; color-favored electroweak-penguin (EW-penguin) amplitude $P_{\rm EW}$; and color-suppressed EW-penguin amplitude $P_{\rm EW}^C$.

2. Weak annihilation amplitudes: $W$-exchange amplitude $E$; $W$-annihilation amplitude $A$; QCD-penguin exchange amplitude $\PE$; QCD-penguin annihilation amplitude $\PA$; EW-penguin exchange amplitude $P\!E_{\rm EW}$; and EW-penguin annihilation amplitude $P\!A_{\rm EW}$.

In this approach, the topological diagrams are classified according to the topologies in the flavor flow of weak decay diagrams, with all strong interaction effects included implicitly in all possible ways.  Therefore, analyses of topological graphs can provide valuable information on final-state interactions.

\subsection{Topological amplitudes}

The topological amplitudes $T,C,E,A$ are extracted from the Cabibbo-favored (CF) $D\to P\!P$ decays~\cite{PDG} to be (in units of $10^{-6}$ GeV) 
\be \label{eq:PP1}
&& T=3.113\pm0.011, \qquad\qquad\qquad\quad
C=(2.767\pm0.029)\,e^{-i(151.3\pm0.3)^\circ}, \non \\
&&  E=(1.48\pm0.04)\,e^{i(120.9\pm0.4)^\circ},
\qquad  A=(0.55\pm0.03)\,e^{i(23^{+~7}_{-10})^\circ}
\en
for $\phi=43.5^\circ$~\cite{Aaij:2014jna}, where $\phi$ is the $\eta-\eta'$ mixing angle defined in the flavor basis
\be
 \begin{pmatrix} \eta \cr \eta' \end{pmatrix}
 = \begin{pmatrix} \cos\phi & -\sin\phi \cr \sin\phi & \cos\phi\cr \end{pmatrix}
 \begin{pmatrix} \eta_q \cr \eta_s \end{pmatrix},
\en
with $\eta_q={1\over\sqrt{2}}(u\bar u+d\bar d)$ and $\eta_s=s\bar s$. The fitted $\chi^2$ value is $0.135$ per degree of freedom. Comparing with the amplitudes obtained in a previous fit in~\cite{ChengChiang}
\be \label{eq:PP2}
&& T=3.14\pm0.06, \qquad\qquad\qquad\quad
C=(2.61\pm0.08)\,e^{-i(152\pm1)^\circ}, \non \\
&&  E=(1.53^{+0.07}_{-0.08})\,e^{i(122\pm2)^\circ},
\qquad\quad  A=(0.39^{+0.13}_{-0.09})\,e^{i(31^{+20}_{-33})^\circ}
\en
we see that the errors in $T$, $C$, $E$ and $A$ are substantially reduced, especially for the annihilation amplitude $A$, thanks to the improved data precision from 2019 PDG~\cite{PDG}.

We note in passing that since we will only fit to the observed branching fractions, the results will be the same if all the strong phases are subject to a simultaneous sign flip.  Throughout this paper, we only present one of them.  Presumably, such a degeneracy in strong phases can be resolved by measurements of sufficiently many \CP asymmetries.

One of the most important moral lessons we have learnt from this approach is that all
the topological amplitudes except the tree amplitude $T$  given in Eq.~(\ref{eq:PP1})
are dominated by nonfactorizable long-distance effects.
For example, in the na{\"i}ve factorization approach, the topological amplitudes $T$ and $C$ in
CF $D\to \bar K\pi$ decays have the expressions
\be \label{eq:T,C}
 T &=& {G_F\over
 \sqrt{2}}a_1(\ov K\pi)\,f_\pi(m_D^2-m_K^2)F_0^{DK}(m_\pi^2), \non \\
 C &=& {G_F\over
 \sqrt{2}}a_2(\ov K\pi)\,f_K(m_D^2-m_\pi^2)F_0^{D\pi}(m_K^2),
\en
with $a_1=c_1+c_2/3$ and $a_2=c_2+c_1/3$. It turns out that $a_1(\ov K\pi)\approx 1.22$ and $a_2(\ov K\pi)\approx 0.82e^{-i(151)^\circ}$~\cite{ChengChiang} extracted from the experimental values of $T$ and $C$ given in Eq.~(\ref{eq:PP1}) and the phenomenological model for the $D$ to $K$ and $\pi$ transition form factors. Since $c_1(m_c)\approx 1.274$ and $c_2(m_c)\approx -0.529$, it is evident that $a_1=c_1+c_2/3\approx 1.09$ is close to $a_1(\ov K\pi)$, while $a_2=c_2+c_1/3\approx -0.11$ expected from na{\"i}ve factorization is far off from $a_2(\ov K\pi)$, including its size and phase. This implies that the short-distance contribution to $C$ is very suppressed relative to the long-distance one. In the topological approach, the long-distance color-suppressed $C$ is induced from the color-allowed $T$ through final-state rescattering with quark exchange.  The nontrivial relative phase between $C$ and $T$ indicates that final-state interactions (FSI's) via quark exchange are responsible for this.

Likewise, short-distance weak annihilation diagrams are helicity suppressed, whereas data imply large sizes of them. This is because they receive large $1/m_c$ power corrections from FSI's and large nonfactorizable contributions for $a_2$. For example, the topological amplitude $E$ receives contributions from the tree amplitude $T$ via final-state rescattering with nearby resonance effects.  The large magnitude and phase of weak annihilation can be quantitatively and qualitatively understood as elaborated in Refs.~\cite{Zen,Chenga1a2}.

As emphasized in~\cite{Cheng:2012a},
one of the great merits of the topological approach is that the magnitude and the relative strong phase of each individual topological tree amplitude in charm decays can be extracted from the data. Consequently, direct \CP asymmetries in charmed meson decays induced at the tree level can be reliably estimated as we shall discuss in Sec.~\ref{sec:tree_induced_CPV}.

\subsection{Flavor SU(3) symmetry breaking}

Using the topological amplitudes in Eq.~(\ref{eq:PP1}) extracted from the CF modes, we can predict the rates for the SCS decays (see the second column of Table~\ref{tab:BFpp} below). It is known that there exists significant SU(3) breaking in some of the SCS modes from the flavor SU(3) symmetry limit.
For example, the rate of  $D^0\to K^+K^-$  is larger than that of $D^0\to \pi^+\pi^-$ by a factor of $2.8$~\cite{PDG}, while the magnitudes of their decay amplitudes should be the same in the SU(3) limit. This is a long-standing puzzle since SU(3) symmetry is expected to be broken roughly at the level of 30\%. Also, the decay $D^0\to K^0\ov K^0$ is almost prohibited in the SU(3) symmetry limit, but the measured branching fraction is of the same order of magnitude as that of $D^0\to \pi^0\pi^0$.

Since SU(3) breaking effects in $D\to P\!P$ decays have been discussed in detail in~\cite{Cheng:2012b}, in this section we will recapitulate the main points and update some of the results.

As stressed in~\cite{Chau:SU(3)}, a most natural way of solving the above-mentioned long-standing puzzles is that the overall seemingly large SU(3) symmetry violation arises from the accumulation of several small and nominal SU(3) breaking effects in the tree amplitudes $T$ and $E$. We will illustrate this point.
Following~\cite{Brod}, we write
\be \label{eq:Amp:D0pipi}
A(D^0\to \pi^+\pi^-) &=&\lambda_d(T+E+P_d+\PE_d+\PA_d)_{\pi\pi}+\lambda_s (P_s+\PE_s+\PA_s)_{\pi\pi} \non \\
&=& {1\over 2}(\lambda_d-\lambda_s)(T+E+\Delta P)_{\pi\pi}-{1\over 2}\lambda_b(T+E+\Sigma P)_{\pi\pi} \ ,
\en
where $\lambda_p\equiv V_{cp}^*V_{up}$ ($p = d,s,b$), the subscript refers to the quark involved in the associated penguin loop, and
\be
\Delta P &\equiv& (P_d+\PE_d+\PA_d)-(P_s+\PE_s+\PA_s), \non \\
\Sigma P &\equiv& (P_d+\PE_d+\PA_d)+(P_s+\PE_s+\PA_s) \ .
\en
Likewise,
\be
A(D^0\to K^+K^-) &=&\lambda_d(P_d+\PE_d+\PA_d)_{_{K\!K}}+\lambda_s (T+E+P_s+\PE_s+\PA_s)_{_{K\!K}} \non \\
&=& {1\over 2}(\lambda_s-\lambda_d)(T+E-\Delta P)_{_{K\!K}}-{1\over 2}\lambda_b(T+E+\Sigma P)_{_{K\!K}} \ .
\en
As far as the rate is concerned, we can neglect the term with the coefficient $\lambda_b$ which is much smaller than $(\lambda_d-\lambda_s)$.
SU(3)-breaking effects in the tree amplitudes $T$ can be estimated in the factorization approach as
\be \label{eq:SU3breakinT}
{T_{_{K\!K}}\over T}={f_K\over f_\pi}\,{F_0^{DK}(m_K^2)\over F_0^{DK}(m_\pi^2)} \ , \qquad\qquad  {T_{\pi\pi}\over T}={m_D^2-m_\pi^2\over m_D^2-m_K^2}\,{F_0^{D\pi}(m_\pi^2)\over F_0^{DK}(m_\pi^2)} \ ,
\en
where $T$ is the tree amplitude in CF $D\to \ov K\pi$ decays given in Eq.~(\ref{eq:T,C}).
Using the form-factor $q^2$ dependence determined experimentally
from Ref.~\cite{CLEO:FF}, we find
\be \label{eq:TKK}
 |T_{_{K\!K}}/T| = 1.269 \ , \qquad \qquad |T_{\pi\pi}/T| = 0.964 \ .
\en

SU(3) symmetry should be also broken in the $W$-exchange amplitudes. This can be seen from the observation of the decay $D^0\to K^0\ov K^0$ whose decay amplitude is given by
\be
A(D^0\to K^0\ov K^0)=\lambda_d(E_d+2\PA_d)+\lambda_s(E_s+2\PA_s) \ ,
\en
with $E_q$ referring to the $W$-exchange amplitude associated with $c\bar u\to q\bar q$ ($q=d,s$).
In the SU(3) limit, the decay amplitude is proportional to $\lambda_b$ and hence its rate is negligibly small, while experimentally $\B(D^0\to K^0\ov K^0)=(0.282\pm0.010)\times 10^{-3}$~\cite{PDG}. This implies sizable SU(3) symmetry violation in the $W$-exchange and QCD-penguin annihilation amplitudes.
Neglecting $P\!A$ and $\lambda_b$ terms and assuming that the $T$ and $E$ amplitudes are responsible for the SU(3) symmetry breaking,
we can fix the SU(3) breaking effects in the $W$-exchange amplitudes from the following four $D^0$ decay modes: $K^+K^-$, $\pi^+\pi^-$, $\pi^0\pi^0$ and $K^0\ov K^0$~\cite{Cheng:2012b}. A fit to the data yields two possible solutions:
\be \label{eq:EdEs}
{\rm I:} &&  E_d=1.10\, e^{i15.1^\circ}E ~, \qquad E_s=0.62\, e^{-i19.7^\circ}E
\ ; \non \\
{\rm II:} &&  E_d=1.10\, e^{i15.1^\circ}E ~, \qquad E_s=1.42\, e^{-i13.5^\circ}E
\ .
\en
The corresponding $\chi^2$ vanishes as these two solutions can be obtained exactly.

If the SU(3)-breaking effects in the $T$ and $C$ topologies are ignored, we find that $\chi^2$ will become very large, of order 340. This is understandable because the large rate disparity between $K^+K^-$ and $\pi^+\pi^-$ cannot rely solely on the nominal SU(3) breaking in the tree or $W$-exchange amplitudes.
When considering SU(3)-breaking effects in $T$, we find that $\B(D^0\to\pi^+\pi^-)$ is reduced slightly from $2.27$ (in units of $10^{-3}$) to 2.11, while
$\B(D^0\to K^+K^-)$ is increased substantially from $1.91$ to 3.15 (see Eq.~(\ref{eq:TKK})). When $E$ is replaced by $E_d=1.10 e^{i15^\circ}E$ in the amplitude of $D^0\to \pi^+\pi^-$, the magnitude of $(0.96T+E_d)$ in $A(D^0\to \pi^+\pi^-)$ becomes smaller than that of $(0.96T+E)$ as the phase of $E$ is about $121^\circ$, so that $\B(D^0\to\pi^+\pi^-)$ is decreased further from 2.11 to 1.47\,.
Likewise, with $E$ being replaced by $E_s=0.62e^{-i20^\circ}E$ or $E_s=1.42e^{-i14^\circ}E$ in the  amplitude of $D^0\to K^+K^-$,  the magnitude of $(1.27T+E_s)$ is enhanced relative to $(1.27T+E)$. It follows that $\B(D^0\to K^+K^-)$ is increased further from 3.15 to 4.03 or 4.05\,.
This shows that the seemingly large SU(3) symmetry violation in $\Gamma(D^0\to K^+K^-)$ and $\Gamma(D^0\to \pi^+\pi^-)$ simply follows from the accumulation of several smaller and nominal SU(3) breaking effects in the tree amplitudes $T$ and $E$.


\begin{table}
\caption{Topological amplitudes for singly Cabibbo-suppressed decays of charmed mesons to two pseudoscalar mesons where flavor SU(3) symmetry breaking effects are included. Summation over $p=d,~s$ is understood.
  \label{tab:PPSCSamp}}
  \footnotesize{
\medskip
\begin{ruledtabular}
\begin{tabular}{l l l }
 & Mode & Representation
      \\
\hline
$D^0$
  & $\pi^+ \pi^-$ & $\lambda_d(0.96T+E_d)+\lambda_p(P_p+\PE_p+\PA_p)$  \\
  & $\pi^0 \pi^0$ & ${1\over \sqrt{2}}\lambda_d(-0.78C+E_d)+{1\over\sqrt{2}}\lambda_p(P_p+\PE_p+\PA_p)$ \\
  & $\pi^0 \eta $ & $-\lambda_d (E_d)\cos\phi-{1\over\sqrt{2}}\lambda_s(1.28 C)\sin\phi+\lambda_p(P_p+\PE_p)\cos\phi$ \\
  & $\pi^0 \eta' $ & $-\lambda_d (E_d)\sin\phi+{1\over\sqrt{2}}\lambda_s (1.28C)\cos\phi+\lambda_p(P_p+\PE_p)\sin\phi$ \\
  & $\eta\eta $ & ${1\over\sqrt{2}}\lambda_d(0.78 C+E_d)\cos^2\phi+\lambda_s(-{1\over 2}1.08C\sin 2\phi+\sqrt{2}\,E_s\sin^2\phi)$+${1\over\sqrt{2}}\lambda_p(P_p+\PE_p+\PA_p)\cos^2\phi$ \\
  & $\eta\eta' $ & ${1\over 2}\lambda_d(0.78 C+E_d)\sin 2\phi+\lambda_s({1\over \sqrt{2}}1.08C\cos 2\phi-E_s\sin 2\phi)$ +${1\over 2}\lambda_p(P_p+\PE_p+\PA_p)\sin 2\phi$ \\
  & $K^+ K^{-}$ & $\lambda_s(1.27T+E_s)+\lambda_p(P_p+\PE_p+\PA_p)$
     \\
  & $K^0 \ol{K}^{0}$ & $\lambda_d (E_d)+\lambda_s (E_s)+2\lambda_p (\PA_p)$  \\
\hline
$D^+$
  & $\pi^+ \pi^0$ & ${1\over\sqrt{2}}\lambda_d(0.97T+0.78C)$
     \\
  & $\pi^+ \eta $ & ${1\over\sqrt{2}}\lambda_d(0.82T+0.93C+1.19A)\cos\phi-\lambda_s(1.28C)\sin\phi+\sqrt{2}
  \lambda_p(P_p+\PE_p)\cos\phi$
      \\
  & $\pi^+ \eta' $ & ${1\over\sqrt{2}}\lambda_d(0.82T+0.93C+1.61A)\sin\phi+\lambda_s(1.28C)\cos\phi+\sqrt{2}\lambda_p(P_p+\PE_p)\sin\phi$
     \\
  & $K^+ \ol{K}^{0}$ & $\lambda_d (0.85A)+\lambda_s(1.28T)+\lambda_p(P_p+\PE_p)$
     \\
\hline
$D_s^+$
  & $\pi^+ K^{0}$ & $\lambda_d(1.00T)+\lambda_s (0.84A)+\lambda_p(P_p+\PE_p)$
     \\
  & $\pi^0 K^{+}$ & ${1\over\sqrt{2}}[-\lambda_d(0.81C)+\lambda_s (0.84A)+\lambda_p(P_p+\PE_p)]$
     \\
  & $K^{+}\eta$ & $\frac{1}{\sqrt{2}}\lambda_p[0.92C\delta_{pd} + 1.14A\delta_{ps}
  + P_p+P\!E_p ]\cos\phi-\lambda_p[(1.31T+1.27C+1.14A)\delta_{ps} + P_p+P\!E_p]\sin\phi$
  \\
  & $K^{+}\eta' $ & $\frac{1}{\sqrt{2}}\lambda_p[0.92C\delta_{pd} +1.14 A\delta_{ps}
  + P_p+P\!E_p]\sin\phi+\lambda_p[(1.31T+1.27C+1.14A)\delta_{ps} + P_p+P\!E_p] \cos\phi$
   \\
\end{tabular}
\end{ruledtabular} }
\end{table}
%


At the hadron level, flavor SU(3) breaking due to the strange and light quark differences will manifest in the decay constants, form factors, wave functions and hadron masses, etc. That is how we evaluate the SU(3)-breaking effect in the $T$ amplitude via Eq.~(\ref{eq:SU3breakinT}). Since the $W$-exchange is governed by long-distance effects, we do not know how to estimate its SU(3) symmetry violation. Hence, we rely on the four modes: $K^+K^-$, $\pi^+\pi^-$, $\pi^0\pi^0$ and $K^0\ov K^0$ to extract $E_d$ and $E_s$.

Different mechanisms have been proposed in the literature for explaining the large rate difference between $D^0\to\pi^+\pi^-$ and $D^0\to K^+K^-$. For example, it has been argued that $\Delta P$ dominated by the difference of $s$- and $d$-quark penguin contractions of 4-quark tree operators is responsible for the large SU(3) breaking in $K^+K^-$ and $\pi^+\pi^-$ modes ~\cite{Brod}.
However, this requires that $|\Delta P/T|\sim 0.5$\,. This mechanism demands a large penguin which is comparable or even larger than $T$. Moreover, it requires a large difference between $s$- and $d$-quark penguin contractions. In Sec.~\ref{sec:penguin_induced_CPV}, we shall see that  $|\Delta P/T|$ is estimated to be of order 0.01 for the short-distance $\Delta P$. Because of the smallness of $\Delta P$, we need to rely on SU(3) violation in both $T$ and $E$ amplitudes to explain the large disparity in the rates of $D^0\to K^+K^-$ and $\pi^+\pi^-$.

Another scenario in which the dominant source of SU(3) breaking lies in final-state interactions was advocated recently in~\cite{Buccella:2019}. To fit the data, several large strong phases such as $\delta_0,\delta_1$ and $\delta_{1/2}$ from final-state interactions are needed~\cite{Buccella:2019}.  They deviate substantially from the SU(3) limit, namely, $\delta_0=\delta_1=\delta_{1/2}$.



\begin{table}
\caption{Branching fractions (in units of $10^{-3}$) of singly Cabibbo-suppressed $D\to PP$ decays.
The column denoted by $\B_{_{\rm SU(3)}}$ shows the predictions based on our best-fitted results in Eq.~(\ref{eq:PP1}) with exact flavor SU(3) symmetry, while SU(3) symmetry breaking effects are taken into account in the column denoted by $\B_{_{\rm SU(3)\!-\!breaking}}$. The first (second) entry in $D^0\to \eta\eta$, $\eta\eta'$, $K^+K^-$ and $K^0\ov K^0$ modes is for Solution I (II) of $E_d$ and $E_s$ in Eq.~(\ref{eq:EdEs}). Experimental results of branching fractions are taken from PDG~\cite{PDG}.
  \label{tab:BFpp}}
\bigskip
\begin{tabular}{l c c c  } \hline\hline
Decay Mode & $\B_{_{\rm SU(3)}}$ & $\B_{_{\rm SU(3)\!-\!breaking}}$ & $\B_{\rm expt}$  \\
 \hline
$D^0\to \pi^+ \pi^-$~~~~ & ~~$2.28\pm0.02$~~ & ~~$1.47\pm0.02$ & ~~$1.455\pm0.024$  \\
$D^0\to\pi^0 \pi^0$  & $1.50\pm0.03$ & ~~$0.82\pm0.02$ & ~~$0.826\pm0.025$  \\
$D^0\to \pi^0 \eta $  & $0.83\pm0.02$ & ~~$0.92\pm0.02$ & ~~$0.63\pm0.06$  \\
$D^0\to \pi^0 \eta' $  & $0.75\pm0.02$ & ~~$1.36\pm0.03$ & ~~$0.92\pm0.10$  \\
$D^0\to \eta\eta $ & $1.52\pm0.03$ & ~~$1.82\pm0.04$ & ~~$2.11\pm0.19$  \\
                   & $1.52\pm0.03$ & ~~$2.11\pm0.04$ & \\
$D^0\to \eta\eta' $  & $1.28\pm0.05$ & ~~$0.69\pm0.03$ & ~~$1.01\pm0.19$  \\
                     & $1.28\pm0.05$ & ~~$1.63\pm0.08$ & \\
$D^0\to K^+ K^{-}$   & $1.91\pm0.02$ & ~~$4.03\pm0.03$ & ~~$4.08\pm0.06$   \\
                     & $1.91\pm0.02$ & ~~$4.05\pm0.05$ &  \\
$D^0\to K_S {K}_{S}$  & 0 & ~~$0.141\pm0.007$ & ~~$0.141\pm0.005$ \\
                         & 0 & ~~$0.141\pm0.007$ & \\
$D^+\to \pi^+ \pi^0$   & $0.89\pm0.02$ & ~~$0.93\pm0.02$ & ~~$1.247\pm0.033$   \\
$D^+\to \pi^+ \eta $  & $1.90\pm0.16$ & ~~$4.08\pm0.16$ & ~~$3.77\pm0.09$  \\
$D^+\to\pi^+ \eta' $  & $4.21\pm0.12$ & ~~$4.69\pm0.08$ & ~~$4.97\pm0.19$   \\
$D^+\to K^+ K_S$  & $2.29\pm0.09$ & ~~$4.25\pm0.10$ & ~~$3.04\pm0.09$   \\
$D_s^+\to \pi^+ K_S$  & $1.20\pm0.04$ & ~~$1.27\pm0.04$ & ~~$1.22\pm0.06$\\
$D_s^+\to \pi^0 K^{+}$  & $0.86\pm0.04$ & ~~$0.56\pm0.02$ & ~~$0.63\pm0.21$  \\
$D_s^+\to K^{+}\eta$   & $0.91\pm0.03$ & ~~$0.86\pm0.03$ & ~~$1.77\pm0.35$  \\
$D_s^+\to K^{+}\eta' $  & $1.23\pm0.06$ & ~~$1.49\pm0.08$ & ~~$1.8\pm0.6$  \\
\hline\hline
\end{tabular}
\end{table}
%


SU(3) breaking effects in the topological amplitudes for SCS $D\to PP$ decays are summarized in Table~\ref{tab:PPSCSamp}. For simplicity, flavor-singlet QCD penguin, flavor-singlet weak annihilation and electroweak penguin annihilation amplitudes have been neglected in subsequent numerical analyses. The reader is referred to Refs.~\cite{ChengChiang,Cheng:2012b} in which we have illustrated SU(3) breaking effects in some selective SCS modes.
The predicted and measured branching fractions are given in Table~\ref{tab:BFpp}.\footnote{Throughout this paper, predictions are made by sampling $10^4$ points in the parameter space, assuming that each of the parameters has a Gaussian distribution with the corresponding central value and symmetrized standard deviation.  Then the predicted values are the mean and standard deviation of data computed using the $10^4$ points.
\label{ftnt}}
While the agreement with experiment is improved for most of the SCS modes after taking into account SU(3) breaking effects in decay amplitudes, there are a few exceptions. For example, the predicted rate for $D^0\to \pi^0\eta^{(')}$ becomes slightly worse compared to the prediction based on SU(3) symmetry even though $D^+\to\pi^+\eta^{(')}$ works better in the presence of SU(3) breaking.

\subsection{Penguin amplitudes in QCD factorization \label{sec:penguin}}

Although the topological tree amplitudes $T,C,E$ and $A$ for hadronic $D$ decays can be extracted from the data, information on penguin amplitudes (QCD penguin, penguin annihilation, etc.) is still needed in order to estimate \CP violation in the SCS decays. To calculate the penguin contributions, we start from the short-distance effective Hamiltonian
\be
{\cal H}_{\rm eff}={G_F\over\sqrt{2}}\left[\sum_{p=d,s}\lambda_p(c_1O_1^p+c_2O_2^p+c_{8g}O_{8g})
-\lambda_b\sum_{i=3}^{6}c_iO_i\right] \ ,
\en
where
\be
&& O_1^p=(\bar pc)_{_{V-A}}(\bar up)_\vma, \qquad\qquad\quad O_2^p=(\bar p_\alpha c_\beta)_{_{V-A}}(\bar u_\beta p_\alpha)_\vma, \non \\
&& O_{3(5)}=(\bar uc)\vma\sum_q(\bar qq)_{_{V\mp A}}, \qquad~
O_{4(6)}=(\bar u_\alpha c_\beta)_\vma\sum_q (\bar q_\beta q_\alpha)_{_{V\mp A}},  \non \\
&& O_{8g}=-{g_s\over 8\pi^2}m_c\,\bar u\sigma_{\mu\nu}(1+\gamma_5)G^{\mu\nu}c \ ,
\en
with $O_3$--$O_6$ being the QCD penguin operators and $(\bar q_1q_2)_{_{V\pm A}}\equiv\bar q_1\gamma_\mu(1\pm \gamma_5)q_2$.
We shall work in the QCD factorization (QCDF) approach~\cite{BBNS99,BN} to evaluate the hadronic matrix elements, but keep in mind that we employ this approach simply for a crude estimate of the penguin amplitudes because the charm quark mass is not heavy enough and $1/m_c$ power corrections are so large that a sensible heavy quark expansion is not allowed.

Let us first consider the penguin amplitudes in $D\to P_1P_2$ decays
\be \label{eq:PinPP}
P^p_{P_1P_2} &=& {G_F \over \sqrt{2}} [a_4^p(P_1P_2)+r_\chi^{P_2} a^p_6(P_1P_2)]f_{P_2} (m_{D}^2 -m^2_{P_1}) ~F_0^{D P_1} (m_{P_2}^2) \ , \non \\
 P\!E_{P_1P_2}^p
 &=& {G_F \over \sqrt{2}}\,
   (f_D f_{P_1} f_{P_2})\left[ b_3^p \right]_{P_1 P_2 } ~,  \\
 P\!A_{P_1P_2}^p
 &=& {G_F \over \sqrt{2}} \,
   ( f_D f_{P_1} f_{P_2})\left[ b_4^p \right]_{P_1 P_2 } ~, \non
\en
where $p=d,s$ and
\begin{eqnarray}
 r_\chi^P(\mu) = {2m_P^2 \over m_c(\mu)(m_2+m_1)(\mu)}
\end{eqnarray}
is a chiral factor.
Here we have followed the conventional Bauer-Stech-Wirbel definition for the form factor $F_{0}^{DP}$~\cite{BSW}. The explicit expressions of the flavor operators $a_4^p$ and $a_6^p$ will be given in Eq.~(\ref{eq:penguinWC}) below.
The annihilation operators $b_{3,4}^p$ are given by
\be \label{eq:bi}
b_3^p &=& {C_F\over N_c^2}\left[c_3A_1^i+c_5(A_3^i+A_3^f)+N_cc_6 A_3^f\right], \non \\
b_4^p &=& {C_F\over N_c^2}\left[c_4A_1^i+c_6A_2^i\right],
\en
where the annihilation amplitudes $A_{1,2,3}^{i,f}$ are defined in Ref.~\cite{BN}.

In practical calculations of QCDF, the superscript `$p$' can be omitted for $a_3,a_5,b_3$ and $b_4$. Hence, we have $\PE^s=\PE^d$, for instance. For $a_4^p$ and $a_6^p$, the terms dictating the `$p$' dependence are $G_{M_2}(s_p)$ and $\hat G_{M_2}(s_p)$, respectively, defined in Eq.~(\ref{eq:G}) below.

\section{Direct \CP violation in $D\to P\!P$ decays \label{sec:DCPV}}

In Ref.~\cite{Cheng:2012b}, we have discussed direct \CP violation in $D\to P\!P$ decays.
Here we will update and improve the results.
For example, we will discuss the issue of end-point divergences with the penguin-exchange and penguin-annihilation amplitudes. We will also consider the uncertainties connected with
long-distance contribution to the penguin-exchange amplitude.
We shall keep some necessary formula presented in~\cite{Cheng:2012b} for ensuing discussions.

\subsection{Tree-level \CP violation}  \label{sec:tree_induced_CPV}

Direct \CP asymmetry in hadronic charm decays defined by
\be
a_{CP}^{\rm dir}(f)={\Gamma(D\to f)-\Gamma(\overline D\to \bar f)\over \Gamma(D\to f)+\Gamma(\overline D\to \bar f)}
\en
can occur even at the tree level~\cite{Cheng1984}. As stressed in~\cite{Cheng:2012a,Cheng:2012b}, the estimate of the tree-level \CP violation  $a_{\rm dir}^{\rm (tree)}$ should be trustworthy since
the magnitude and the relative strong phase of each individual topological tree amplitude in charm decays can be extracted from the data. The predicted tree-level \CP asymmetries for SCS modes are shown in Table~\ref{tab:CPVpp}. We see that larger \CP asymmetries can be achieved in those decay modes with interference between $T$ and $C$ or $C$ and $E$. For example, $a_{\rm dir}^{({\rm tree})}$ is of order $0.78\times 10^{-3}$ for $D^0\to \pi^0\eta$ and $-0.75\times 10^{-3}$ for $D_s^+\to K^+\eta$.

Direct \CP violation in $D^0\to K_SK_S$ is given by
\be \label{eq:acpKK}
a_{\rm dir}^{({\rm tree})}(D^0\to K_SK_S)= {2{\rm Im}(\lambda_d\lambda_s^*)\over |\lambda_d|^2}\,
{{\rm Im}(E_d^*E_s)\over |E_d-E_s|^2}= 1.3\times 10^{-3} {|E_dE_s|\over |E_d-E_s|^2}\sin\delta_{ds} \ ,
\en
where $\delta_{ds}$ is the strong phase of $E_s$ relative to $E_d$. From the two solutions of $E_d$ and $E_s$ given in Eq.~(\ref{eq:EdEs}), we find
\footnote{In our previous work~\cite{Cheng:2012b}, we obtained $a_{\rm dir}^{({\rm tree})}(D^0\to K_SK_S)=-0.7\times 10^{-3}$ for Solution I and $-1.7\times 10^{-3}$ for Solution II.}
\be \label{eq:KSKS}
a_{\rm dir}^{({\rm tree})}(D^0\to K_SK_S)=\left\{
\begin{array}{cl}
    -1.05\times 10^{-3}
      & \quad \mbox{Solution~I} \ , \\
    -1.99\times 10^{-3}
      & \quad \mbox{Solution~II} \ .
    \end{array}\right.
\en
For comparison, various predictions available in the literature are discussed here.
$a_{\rm dir}^{({\rm tree})}(K_SK_S)=1.11\times 10^{-3}$ was predicted in~\cite{Li:2012}.
It ranges in $(0.38-0.43)\times 10^{-3}$ according to~\cite{Buccella:2019} (see also the last column of Table~\ref{tab:CPVpp}).  Both predictions are of the opposite sign from ours. As explained in~\cite{Cheng:2012b}, the positive sign of $a_{\rm dir}^{({\rm tree})}(K_SK_S)$ given in~\cite{Li:2012} can be traced back to the phase of the $W$-exchange amplitude. In our case, the $W$-exchange amplitude is always in the second quadrant, while it lies in the third quadrant in~\cite{Li:2012} due to a sign flip.
As noticed in passing, all the strong phases extracted from a fit to branching fractions are equivalent to those with a simultaneous sign flip. This explains why the strong phases of $C$ and $E$ in~\cite{Li:2012} are simultaneously opposite to ours in sign, and the sign difference between this work and~\cite{Li:2012} for $a_{\rm dir}^{({\rm tree})}(K_SK_S)$. A measurement of $a_{CP}^{\rm dir}(D^0\to K_SK_S)$ will resolve the discrete phase ambiguity. If it is measured to be negative as predicted by us, then the $W$-exchange amplitude should be in the second quadrant.

In~\cite{Hiller}, the direct \CP violation in $D^0\to K_SK_S$ was connected to that of $D^0\to K^+K^-$ via the relation
\be
{ a^{\rm dir}_{CP}(D^0\to K_SK_S)\over a^{\rm dir}_{CP}(D^0\to K^+K^-)}\sim \sqrt{ {\B(D^0\to K^+K^-)\over 2\B(D^0\to K_SK_S)} }.
\en
Taking $a^{\rm dir}_{CP}(D^0\to K^+K^-)$ to be $(-0.48\pm0.09)\times 10^{-3}$ from  Table~\ref{tab:CPVpp} and the measured branching fractions,  the obtained result  $a^{\rm dir}_{CP}(D^0\to K_SK_S)\approx -1.8\times 10^{-3}$ is in agreement in magnitude and sign with ours.
$a^{\rm dir}_{CP}(D^0 \to K_SK_S)$ was estimated to be 0.6\% in~\cite{Brod}, while an upper bound
$|a_{\rm dir}(D^0\to K_SK_S)|\leq 1.1\%$ was set in~\cite{Nierste}.

The current experimental measurements are
\be
a^{\rm dir}_{CP}(K_SK_S)=
\begin{cases}
	(-2.9\pm5.2\pm2.2)\% & \mbox{LHCb~\cite{LHCb:KSKS2015}},  \\
    (4.3\pm3.4\pm1.0)\% & \mbox{LHCb~\cite{LHCb:KSKS2018}},  \\
    (-0.02\pm1.53\pm0.17)\%  & \mbox{Belle~\cite{Belle:KSKS}}.
\end{cases}
\en
Since LHCb has measured $\Delta A_{CP}$ to the accuracy of $10^{-3}$,
it is conceivable that an observation of \CP violation in the decay $D^0\to K_SK_S$ will be feasible in the near future.

\begin{table}
\footnotesize{
\caption{Direct \CP asymmetries (in units of $10^{-3}$) of $D\to PP$ decays, where $a_{\rm dir}^{({\rm tree})}$ denotes \CP asymmetry arising from purely tree amplitudes.
The superscript (t+p) denotes tree plus QCD-penguin amplitudes, (t+pa) for tree plus weak penguin-annihilation ($P\!E$ and $P\!A$) amplitudes and ``tot" for the total amplitude.
The first (second) entry in $D^0\to \eta\eta$, $\eta\eta'$, $K^+K^-$ and $K_SK_S$ is for Solution I (II) of $E_d$ and $E_s$ [Eq.~(\ref{eq:EdEs})].
For QCD-penguin exchange $\PE$, we assume that it is similar to the topological $E$ amplitude [see Eq.~(\ref{eq:PE})].   For comparison, The predicted results of  $a_{\rm dir}^{({\rm tot})}$ in~\cite{Buccella:2019} for both the negative (former) and positive (latter) solutions for the phase $\delta_i$  are also presented.
  \label{tab:CPVpp}}
\medskip
\begin{ruledtabular}
\begin{tabular}{l c  c c c c c }
Decay Mode &  $a_{\rm dir}^{({\rm tree})}$~ &  $a_{\rm dir}^{({\rm t+p})}$~ &  $a_{\rm dir}^{({\rm t+pa})}$~
     & $a_{\rm dir}^{({\rm tot})}$(This work)~   & $a_{\rm dir}^{({\rm tot})}$\cite{Buccella:2019}~   \\
\hline
$D^0\to \pi^+ \pi^-$ & $0$ & $0.03\pm0.01$ & $0.78\pm0.22$
     & $0.80\pm0.22$ & $1.17\pm0.20~/~1.18\pm0.20$  \\
$D^0\to\pi^0 \pi^0$  & $0$ & $0.27\pm0.01$ & $0.55\pm0.30$
     & $0.82\pm0.30$ & $0.04\pm0.09~/~0.79\pm0.10$  \\
$D^0\to \pi^0 \eta $  &  $0.78\pm0.01$  & $0.48\pm0.01$ & $0.24\pm0.28$   & $-0.05\pm0.28$~~   \\
$D^0\to \pi^0 \eta' $  &  $-0.43\pm0.01$~~  & $-0.56\pm0.01$~~ & $-0.01\pm0.17$   & $-0.15\pm0.17$~~   \\
$D^0\to \eta\eta $ &  $-0.28\pm0.01$~~ & $-0.28\pm0.01$~~ & $-0.51\pm0.07$~~  & $-0.52\pm0.07$~~  \\
& $-0.37\pm0.01$~~ & $-0.44\pm0.01$~~ & $-0.58\pm0.07$~~  & $-0.65\pm0.07$~~   \\
$D^0\to \eta\eta' $  &  $0.51\pm0.00$ & $0.09\pm0.00$ & $0.72\pm0.22$    & $0.29\pm0.21$    \\
 &  $0.46\pm0.01$   & $0.16\pm0.00$ & $0.52\pm0.15$   & $0.22\pm0.15$    \\
$D^0\to K^+ K^{-}$   &  $0$ & $0.08\pm0.00$ & $-0.41\pm0.14$~~
     & $-0.33\pm0.14$~~ & $-0.47\pm0.08 ~/ -0.46\pm0.08$~~   \\
     &  $0$ & $-0.01\pm0.00$~~ & $-0.43\pm0.12$~~
     & $-0.44\pm0.12$~~  \\
$D^0\to K_S {K}_{S}$  & $-1.05$ & $-1.05$ & $-1.05$ & $-1.05$~~ & $0.43\pm0.07 ~/~ 0.38\pm0.07$   \\
& $-1.99$ & $-1.99$ & $-1.99$ & $-1.99$~~ \\
$D^+\to \pi^+ \pi^0$  & 0 & 0 & $0$
     & $0$  \\
$D^+\to \pi^+ \eta $  & $0.37\pm0.02$ & $0.07\pm0.01$ & $-0.34\pm0.22$~~ &  $-0.63\pm0.23$~~     \\
$D^+\to\pi^+ \eta' $  &  $-0.26\pm0.02$~~ & $-0.45\pm0.03$~~ & $0.30\pm0.18$  & $0.11\pm0.18$     \\
$D^+\to K^+ {K}_{S}$~~~  &  $-0.07\pm0.02$~~ & $0.10\pm0.02$ & $-0.46\pm0.18$~~    & $-0.30\pm0.18$~~ & $-0.40\pm0.07 ~/ -0.26\pm0.05$~~   \\
$D_s^+\to \pi^+ K_{S}$  &  $0.09\pm0.03$ & $-0.08\pm0.03$~~ & $0.61\pm0.24$    & $0.42\pm0.24$
& $-0.40\pm0.07 ~/ -0.36\pm0.07$~~\\
$D_s^+\to \pi^0 K^{+}$  & $-0.04\pm0.06$~~ & $-0.02\pm0.04$~~ & $0.89\pm0.27$
     & $0.91 \pm 0.27$ & ~~~$0.48\pm0.06 ~/ -0.03\pm0.04$~~     \\
$D_s^+\to K^{+}\eta$   &  $-0.75\pm0.01$~~ & $-0.92\pm0.02$~~ & $-0.64\pm0.08$~~  & $-0.81\pm0.08$~~     \\
$D_s^+\to K^{+}\eta' $  &  $0.34\pm0.02$ & $0.63\pm0.03$ & $-0.22\pm0.24$~~     & $0.07\pm0.25$~~      \\
\end{tabular}
\end{ruledtabular}
}
\end{table}


\subsection{Penguin-induced \CP violation \label{sec:penguin_induced_CPV}}

Direct \CP violation does not occur at the tree level in $D^0\to K^+K^-$ and $D^0\to\pi^+\pi^-$. In these two decays, the \CP asymmetry arises from the interference between tree and penguin amplitudes. From Eq.~(\ref{eq:Amp:D0pipi}), we obtain
\be \label{eq:pipiacp}
a^{\rm dir}_{CP}(\pi^+\pi^-) &=& {4{\rm Im}[(\lambda_d-\lambda_s)\lambda_b^*]\over |\lambda_d-\lambda_s|^2}\,
{{\rm Im}[(T^*+E^*+\Delta P^*)(T+E+\Delta P+\Sigma P-\Delta P)]_{\pi\pi}\over |T+E+\Delta P|_{\pi\pi}^2 }   \non \\
&\approx& 1.30\times 10^{-3} \left| {P_s+P\!E_s+P\!A_s\over T+E+\Delta P}\right|_{\pi\pi}\sin\delta_{\pi\pi} \ ,
\en
where $\delta_{\pi\pi}$ is the strong phase of $(P_s+P\!E_s+P\!A_s)_{\pi\pi}$ relative to $(T+E+\Delta P)_{\pi\pi}$ and likewise for $a_{CP}^{\rm dir}(K^+K^-)$.
Hence,
\be \label{eq:DCPVdiff}
\Delta a_{\rm CP}^{\rm dir}=-1.30\times 10^{-3} \left( \left| {P_d+P\!E_d+P\!A_d \over T+E-\Delta P}\right|_{_{K\!K}}\sin\delta_{_{K\!K}}+\left| {P_s+P\!E_s+P\!A_s\over T+E+\Delta P}\right|_{\pi\pi}\sin\delta_{\pi\pi}\right),
\en
with $\delta_{_{K\!K}}$ being the strong phase of $(P_d+P\!E_d+P\!A_d)_{_{K\!K}}$ relative to $(T+E-\Delta P)_{_{K\!K}}$.

Using the input parameters for the light-cone distribution amplitudes of light mesons, quark masses and decay constants from Refs.~\cite{CCBud,Laiho} and form factors from Refs.~\cite{ChengChiang,YLWu},
we find to the leading order in $\Lambda_{\rm QCD}/m_b$ in QCDF that
\be \label{eq:Pover T}
&& \left({P_d\over T}\right)_{\pi\pi}=0.226\, e^{-i150^\circ}, \qquad \left({P_s\over T}\right)_{\pi\pi}=0.231\, e^{-i152^\circ}, \qquad
\left({\Delta P\over T}\right)_{\pi\pi}=0.010\, e^{-i35^\circ}, \non \\
&& \left({P_d\over T}\right)_{_{K\!K}}=0.220\, e^{-i150^\circ}, \qquad \left({P_s\over T}\right)_{_{K\!K}}=0.227\, e^{-i152^\circ}, \qquad
\left({\Delta P\over T}\right)_{_{K\!K}}=0.010\, e^{-i35^\circ}.
\en
It is obvious that $\Delta P=P_d-P_s$ arising from the difference in the $d$- and $s$-loop penguin contractions [see Eq.~(\ref{eq:penguinWC})] is very small compared to the tree amplitude.   It is straightforward to show
\be \label{eq:PoverT+E}
\left({P_s\over T+E+\Delta P}\right)_{\pi\pi}=0.32\, e^{i176^\circ}, \qquad
\left({P_d\over T+E-\Delta P}\right)_{_{K\!K}}=
\begin{cases}
0.23\,e^{-i164^\circ} \\
0.23\,e^{i178^\circ}
\end{cases},
\en
for Solutions I and II of $W$-exchange amplitudes $E_d$ and $E_s$ (see Eq.~(\ref{eq:EdEs})). It follows from Eq.~(\ref{eq:DCPVdiff}) that  $a^{\rm dir}_{CP}(\pi^+\pi^-)=0.029\times 10^{-3}$, and
\be
a^{\rm dir}_{CP}(K^+K^-)=
\begin{cases}
0.082\times 10^{-3} \\
-0.010\times 10^{-3}
\end{cases}, \qquad
\Delta a_{CP}^{\rm dir}\approx
\begin{cases}
0.05\times 10^{-3} & \mbox{Solution I}, \\
-0.02\times 10^{-3} & \mbox{Solution II}.
\end{cases}
\en
Evidently,  \CP asymmetries in $D^0\to \pi^+\pi^-,~ K^+K^-$ induced by QCD penguins are very small mainly due to the strong phases $\delta_{\pi\pi}$ and $\delta_{K\!K}$ being not far from $180^\circ$.

So far we have only discussed leading-order QCDF calculations except for the chiral enhanced
penguin contributions, namely, the $a_6$ terms in Eq.~(\ref{eq:PinPP}).  For QCD-penguin power corrections, we shall consider weak penguin annihilation, namely, QCD-penguin exchange $\PE$ and QCD-penguin annihilation $\PA$ which are formally of order $1/m_c$.
However, it is well known that the weak penguin annihilation amplitudes in QCDF derived from Eq.~(\ref{eq:PinPP})
involve troublesome endpoint divergences~\cite{BBNS99,BN}. Hence,
subleading power corrections generally can be studied only
in a phenomenological way. For example, the endpoint divergence is parameterized as \cite{BBNS99,BN}
\be
X_A\equiv \int_0^1 {dx\over 1-x}=\ln\left({m_D\over \Lambda_h}\right)(1+\rho_A e^{i\phi_A}),
\en
with $\Lambda_h$ being a typical hadronic scale of order 500 MeV, and $\rho_A$, $\phi_A$ being unknown real parameters. In hadronic $B$ decays, the values of $\rho_A$ and $\phi_A$ can be obtained from a fit to $B\to P\!P,V\!P$ and $V\!V$ decays~\cite{Cheng:2009}. However, this is not available in charmed meson decays since penguin effects manifest mainly in \CP violation. Therefore, we will not evaluate $\PE$ and $\PA$ in this way in the charm sector. Nevertheless, if we borrow typical values of $\rho_A$ and $\phi_A$ from the $B$ system, we find weak penguin annihilation contributions smaller than QCD penguin; for instance, $(\PE/T)_{\pi\pi}\sim 0.04$ and $(\PA/T)_{\pi\pi}\sim -0.02$. Therefore, it is safe to neglect short-distance contributions to weak penguin annihilation amplitudes.

As pointed out in~\cite{Cheng:2012a}, long-distance contributions to SCS decays, for example, $D^0\to \pi^+\pi^-$, can proceed through the weak decay $D^0\to K^+K^-$ followed by a resonant-like final-state rescattering as depicted in Fig.~2 of~\cite{Cheng:2012a}. It has the same topology as the QCD-penguin exchange topological graph $P\!E$.  Since weak penguin annihilation and FSI's are both of order $1/m_c$ in the heavy quark limit, this means FSI's could play an essential role in charm decays.
Hence, it is plausible to assume that $P\!E$ is of the same order of magnitude as $E$. In~\cite{Cheng:2012a}, we took $(P\!E)^{\rm LD}=1.60\,e^{i 115^\circ}$ (in units of $10^{-6}$ GeV). In this work we will assign by choice the same magnitude and phase as $E$ with 20\% and $30^\circ$ uncertainties, respectively, so that
\be \label{eq:PE}
(P\!E)^{\rm LD} \approx (1.48\pm0.30)\,e^{i (120.9\pm30.0)^\circ} \ .
\en
For simplicity, we shall assume its flavor independence, that is, $(P\!E)^{\rm LD}_d=(P\!E)^{\rm LD}_s$.

Including the long-distance contribution to penguin exchange $P\!E$, we get
\be \label{eq:fullratio}
\left({P_s+P\!E_s^{\rm LD} \over T+E+\Delta P}\right)_{\pi\pi}=0.77\, e^{i114^\circ},\qquad
\left({P_d+P\!E_d^{\rm LD} \over T+E-\Delta P}\right)_{_{K\!K}}=
\begin{cases}
0.45\, e^{i137^\circ} \\ 0.45\,e^{i120^\circ}
\end{cases}.
\en
As shown in Table~\ref{tab:CPVpp}, we see that the predicted \CP violation denoted by $a_{\rm dir}^{(\rm tot)}$ or $a_{\rm dir}^{(\rm tree)}$
is at most of order $10^{-3}$ in the SM.  Specifically, we have
\footnote{Since Eqs. (\ref{eq:fullratio}) and (\ref{eq:pipiacp}) lead to $a_{CP}^{\rm dir}(\pi^+\pi^-) =0.91\times 10^{-3}$ and $a_{CP}^{\rm dir}(K^+K^-) = -0.40\times 10^{-3}$ for Solution I and $-0.51\times 10^{-3}$ for Solution II, the reader may wonder why they are slightly larger in magnitude than the final results presented in Table~\ref{tab:CPVpp}.   Such a difference is related to the fact that the predictions are made, as alluded to in Footnote~\ref{ftnt}, statistically and the fact that \CP asymmetries are not linear in the parameters.}
\be \label{eq:CPinpipi}
a_{CP}^{\rm dir}(\pi^+\pi^-) &=& (0.80\pm0.22)\times 10^{-3}, \\
a_{CP}^{\rm dir}(K^+K^-) &=&
\begin{cases} (-0.33\pm0.14)\times 10^{-3}  & \mbox{Solution I}, \\
(-0.44\pm0.12)\times 10^{-3} & \mbox{Solution II}.
\end{cases}
\en
Theoretical uncertainties are dominated by that of $(P\!E)^{\rm LD}$.
Hence, the \CP asymmetry difference between $D^0\to K^+K^-$ and $D^0\to \pi^+\pi^-$ is given by
\be
\Delta a_{CP}^{\rm dir}=
\begin{cases}
(-1.14\pm0.26)\times 10^{-3} & \mbox{Solution I}, \cr
(-1.25\pm0.25)\times 10^{-3} & \mbox{Solution II}.
\end{cases}
\en
Although our new results of $\Delta a_{CP}^{\rm dir}$ are slightly smaller than the previous ones in~\cite{Cheng:2012b}, they have more realistic estimates of uncertainties and are consistent with the LHCb's new measurement in Eq.~(\ref{eq:LHCb:2019}) within $1\sigma$.  Here we note in passing that the \CP asymmetry predictions are very sensitive to $(P\!E)^{\rm LD}$.  Had we chosen to use the value of $1.60 \times 10^{-6}\,e^{i 121^\circ}$~GeV, as done in~\cite{Cheng:2012a}, $\Delta a_{CP}^{\rm dir}$ would become $(-1.24\pm0.26)\times 10^{-3}$ for Solution I and $(-1.34\pm0.25)\times 10^{-3}$ for Solution II.


\subsection{Comparison with Li et al.~\cite{Li:2012}}

Based on the so-called factorization-assisted topological-amplitude approach, an estimate of $\Delta a_{CP}=-1.00\times 10^{-3}$ in the SM was made in~\cite{Li:2012}. In this work, the topological amplitudes in units of $10^{-6}$ GeV are given by
\footnote{In terms of the notation of~\cite{Li:2012}, $P,P\!E,P\!A$ correspond to $P_C,P_E$ and $P_A$, respectively.}
\be \label{eq:Amp:Li}
(T,E,P,P\!E,P\!A)_{\pi\pi} &=& (2.73, 0.82e^{-i142^\circ}, 0.87e^{i134^\circ}, 0.81e^{i111^\circ}, 0.25e^{-i43^\circ}),  \non \\
(T,E,P,P\!E,P\!A)_{_{K\!K}} &=& (3.65, 1.20e^{-i85^\circ}, 1.21e^{i135^\circ}, 0.87e^{i111^\circ}, 0.45e^{-i5^\circ}).
\en
As a result,
\be
\left({P+P\!E+P\!A \over T+E }\right)_{\pi\pi}=0.66\, e^{i134^\circ}, \qquad
\left({P+P\!E+P\!A \over T+E }\right)_{_{K\!K}}= 0.45\, e^{i131^\circ}.
\en
This leads to the aforementioned value of $\Delta a_{CP}$. For comparison, in our case we have
\be \label{eq:Amps}
(T,E,P,P\!E)_{\pi\pi} &=& (3.00, 1.64e^{i136^\circ}, 0.69e^{-i152^\circ}, 1.48e^{i121^\circ}),  \non \\
(T,E,P,P\!E)_{_{K\!K}} &=& (3.96, 0.93e^{i101^\circ}, 0.88e^{-i150^\circ}, 1.48e^{i121^\circ})~~~{\rm Solution~I},  \\
 &=& (3.96, 2.10e^{i107^\circ}, 0.88e^{-i150^\circ}, 1.48e^{i121^\circ}) ~~~{\rm Solution~II}. \non
\en

There are three crucial differences between this work and~\cite{Li:2012}: (i) the phase of $E$ amplitudes is in the second quadrant in the former while in the third or fourth quadrant in the latter, (ii) the phase of the penguin amplitude $P$ is in the third quadrant in our work while in the second quadrant in~\cite{Li:2012}, and (iii) our $\PE$ amplitude comes from long-distance final-state rescattering as we have neglected short-distance contributions to weak penguin annihilation amplitudes $\PE$ and $\PA$. As discussed in passing, there is a discrete phase ambiguity for the phases of $C,E$ and $A$ topological amplitudes in our analysis. Presumably, a measurement of $a_{CP}^{\rm dir}(D^0\to K_SK_S)$ will resolve the discrete phase ambiguity for the $E$ amplitude. However, the phase of the penguin amplitude is calculated in theory. Let us examine this issue as follows.

Consider the penguin amplitude $P^p_{P_1P_2}$ given in Eq.~(\ref{eq:PinPP}).
Within the framework of QCDF, the
flavor operators $a_{4,6}^p$ are basically the Wilson coefficients in conjunction with short-distance nonfactorizable corrections such as vertex corrections $V_i$, penguin contractions ${\cal P}_i$ and hard spectator interactions $H_i$:
\be \label{eq:penguinWC}
a^p_4(P_1P_2) &=& \left(c_4+{c_3\over N_c}\right)+{c_3\over N_c}\,{C_F\alpha_s\over 4\pi}[V_4(P_2)+{4\pi^2\over N_c}H_4(P_1P_2)]+{\cal P}^p_4(P_2),  \non \\
a^p_6(P_1P_2) &=& \left(c_6+{c_5\over N_c}\right)+{c_5\over N_c}\,{C_F\alpha_s\over 4\pi}[V_6(P_2)+{4\pi^2\over N_c}H_6(P_1P_2)]+{\cal P}^p_6(P_2),
\en
where the explicit expressions of $V_i$ and $H_i$ can be found in~\cite{BN}. The order $\alpha_s$ corrections from penguin contraction read~\cite{BN}
\be \label{eq:P46}
{\cal P}^p_4&=&{C_F\alpha_s\over 4\pi N_c}\Bigg\{ c_1\left[ {4\over 3}{\rm ln}{m_c\over \mu}+{2\over 3}-G_{M_2}(s_p)\right]+c_3\left[ {8\over 3}{\rm ln}{m_c\over \mu}+{4\over 3}-G_{M_2}(s_u)-G_{M_2}(1)\right]  \non \\
&& +(c_4+c_6)\left[{16\over 3}{\rm ln}{m_c\over \mu}-G_{M_2}(s_u)-G_{M_2}(s_d)-G_{M_2}(s_s)-G_{M_2}(1)\right]-2c_{8g}^{\rm eff}\int^1_0 {dx\over 1-x}\Phi_{M_2}(x) \Bigg\} \ ,  \non \\
{\cal P}^p_6&=&{C_F\alpha_s\over 4\pi N_c}\Bigg\{ c_1\left[ {4\over 3}{\rm ln}{m_c\over \mu}+{2\over 3}-\hat G_{M_2}(s_p)\right]+c_3\left[ {8\over 3}{\rm ln}{m_c\over \mu}+{4\over 3}-\hat G_{M_2}(s_u)-\hat G_{M_2}(1)\right]   \\
&& +(c_4+c_6)\left[{16\over 3}{\rm ln}{m_c\over \mu}-\hat G_{M_2}(s_u)-\hat G_{M_2}(s_d)-\hat G_{M_2}(s_s)-\hat G_{M_2}(1)\right] -2c_{8g}^{\rm eff} \Bigg\} \ ,  \non
\en
where $c_{8g}^{\rm eff}=c_{8g}+c_5$,
$s_i=m_i^2/m_c^2$,
\be \label{eq:G}
G_{M_2}(s)=\int_0^1 dx\,G(s,1-x)\Phi_{M_2}(x), \qquad
\hat G_{M_2}(s)=\int_0^1 dx\,G(s,1-x)\Phi_{m_2}(x),
\en
and $G(s,x)=-4\int^1_0 du\,u(1-u){\rm ln}[s-u(1-u)x]$. Here $\Phi_{M_2}$ ($\Phi_{m_2}$) is the twist-2 (-3) light-cone distribution amplitude for the meson $M_2$.

In~\cite{Li:2012}, the flavor operators $a_{4,6}$ and $a_{1,2}$ are taken to be
\be \label{eq:a46:Li}
&& a_1(\mu)=c_1(\mu)+{c_2(\mu)\over N_c}, \qquad a_2(\mu)=c_2(\mu)+c_1(\mu)\left[ {1\over N_c}+\chi_{nf}e^{i\phi}\right], \non \\
&& a_{4,6}(\mu) = c_{4,6}(\mu)+c_{3,5}(\mu)\left[ {1\over N_c}+\chi_{nf}e^{i\phi}\right],
\en
Comparing Eq.~(\ref{eq:a46:Li}) with Eq.~(\ref{eq:penguinWC}), we see that the source of the QCD penguin's strong phase is assumed to be the same as that of $a_2$ in~\cite{Li:2012}, while it arises from nonfactorizable contributions in QCDF. In other words, while we consider the effects of vertex corrections, penguin contractions and hard spectator interactions for the QCD penguin amplitude, these effects are parameterized in~\cite{Li:2012} in terms of $\chi_{nf}$ and $\phi$, which are determined from a global fit to the measured branching fractions. Since the color-suppressed $C$ amplitude in~\cite{Li:2012} is in the second quadrant, so is the penguin amplitude.
This explains the difference between our work and~\cite{Li:2012} for the QCD penguin amplitudes.

\subsection{Comparison with Chala et al.~\cite{Chala:2019fdb}}
Based on the light-cone sum rule calculations of
\be
\left|{P\over T+E}\right|_{\pi\pi}=0.093\pm0.011, \qquad \left|{P\over T+E}\right|_{_{KK}}=0.075\pm0.015,
\en
Khodjamirian and Petrov~\cite{Khodjamirian:2017} argued an upper bound in the SM, $|\Delta a_{CP}^{\rm SM}|\leq (2.0\pm0.3)\times 10^{-4}$. Including higher-twist effects in the operator product expansion for the underlying correlation functions which are expected to be
\be
\left|{P\over T+E}\right|_{\pi\pi}=0.093\pm0.030, \qquad \left|{P\over T+E}\right|_{_{KK}}=0.075\pm0.035,
\en
Chala {\it et al.}~\cite{Chala:2019fdb} claimed a modification of the SM bound, $|\Delta a_{CP}^{\rm SM}|\leq (2.0\pm1.0)\times 10^{-4}$.

This conclusion seems to be very na{\"i}ve. First,
as stated in~\cite{Khodjamirian:2017}, Khodjamirian and Petrov have neglected the contributions from the penguin operators $O_{i=3,\cdots,6,8g}$ due to their small Wilson coefficients. This means they only considered the penguin contraction from the tree operators $O_{1,2}$. Consequently,
\be
a_4^p=a_6^p \approx {C_F\alpha_s\over 4\pi N_c}c_1\left[ {4\over 3}{\rm ln}{m_c\over \mu}+{2\over 3}-G_{M_2}(s_p)\right].
\en
Secondly, penguin-exchange and penguin-annihilation contributions have not been considered, not mentioning the possible final-state resattering effect on $\PE$. They play an essential role in understanding the LHCb measurement of $\Delta a_{CP}$. Otherwise, it is premature to claim the necessity of New Physics in this regard.

\section{$D \to V\!P$ Decays \label{sec:vp}}

In the treatment of $D \to V\!P$ decays, we continue to use the same topological diagram notation as in the $P\!P$ decays, except that a subscript of $V$ or $P$ is attached to the flavor amplitudes and the associated strong phases to denote whether the spectator quark in the charmed meson ends up in the vector or pseudoscalar meson in the final state.  The $V$-type and $P$-type parameters are completely independent {\it a priori}, though certain relations can be established under the factorization assumption.

\subsection{Topological amplitudes}

The partial decay width of the $D$ meson into a vector and a pseudoscalar mesons are usually expressed in two different ways:
\begin{equation}
\Gamma(D\to VP)=\frac{p_c^3}{8\pi m_D^2}|\tilde{{\cal M}}|^2 ~,
\label{decaywidthA}
\end{equation}
and
\begin{equation}
\Gamma(D\to VP)
=\frac{p_c^3}{8\pi m_V^2}|{{\cal  M}}|^2 ~.
\label{decaywidthC}
\end{equation}
Even though both formulas have the same cubic power dependence on $p_c$ (as required for a P-wave configuration), a main difference resides in the fact that the latter has incorporated an additional SU(3)-breaking factor for the phase space, resulting from the sum of possible polarizations of the vector meson in the final state.

\begin{table}[t]
\caption{Fit results using Eq.~(\ref{decaywidthC}) and $\phi=43.5^\circ$.  The amplitude sizes are quoted in units of $10^{-6}$ and the strong phases in units of degrees.
}
\label{tab:CFVPB}
\medskip
\footnotesize{
\begin{ruledtabular}
\begin{tabular}{c c c c c c c}
& (S1) & (S2) & (S3) & (S4) & (S5) & (S6)
\\
\hline
$|T_V|$        & $2.18^{+0.06}_{-0.07}$ & $2.18^{+0.06}_{-0.07}$ & $2.17\pm0.06$ & $2.19^{+0.06}_{-0.07}$
               & $2.18^{+0.06}_{-0.07}$ & $2.18\pm0.06$
\\
$|T_P|$        & $3.41\pm0.06$ & $3.36\pm0.06$ & $3.51\pm0.06$ & $3.48\pm0.06$ & $3.50\pm0.06$ & $3.39\pm0.06$
\\
$\delta_{T_P}$ & $69\pm3$ & $286\pm3$ & $40^{+3}_{-4}$ & $307^{+4}_{-3}$ & $79^{+3}_{-4}$ & $12\pm3$
\\
$|C_V|$        & $1.76\pm0.04$ & $1.76\pm0.04$ & $1.74\pm0.04$ & $1.75\pm0.04$ & $1.74\pm0.04$ & $1.76\pm0.04$
\\
$\delta_{C_V}$ & $278\pm3$ & $76\pm3$ & $195^{+4}_{-3}$ & $152^{+3}_{-4}$ & $235^{+4}_{-3}$ & $221\pm3$
\\
$|C_P|$        & $2.10\pm0.03$ & $2.07\pm0.03$ & $2.04\pm0.03$ & $2.14\pm0.03$ & $2.07\pm0.03$ & $2.07\pm0.03$
\\
$\delta_{C_P}$ & $201\pm1$ & $201\pm1$ & $201\pm1$ & $159\pm1$ & $159\pm1$ & $201\pm1$
\\
$|E_V|$        & $0.27\pm0.04$ & $0.26\pm0.04$ & $0.40\pm0.06$ & $0.33\pm0.05$ & $0.38\pm0.05$ &$0.26\pm0.04$
\\
$\delta_{E_V}$ & $260^{+50}_{-20}$ & $69^{+46}_{-21}$ & $245^{+8}_{-9}$ & $113^{+14}_{-11}$ & $282^{+8}_{-10}$
               & $224^{+22}_{-40}$
\\
$|E_P|$        & $1.66^{+0.05}_{-0.06}$ & $1.66^{+0.05}_{-0.06}$ & $1.66\pm0.05$
               & $1.66^{+0.05}_{-0.06}$ & $1.66\pm0.05$ & $1.66^{+0.05}_{-0.06}$
\\
$\delta_{E_P}$ & $108\pm3$ & $108\pm3$ & $107\pm3$ & $251\pm3$ & $252\pm3$ & $108\pm3$
\\
$|A_V|$        & $0.19\pm0.02$ & $0.20\pm0.03$ & $0.22\pm0.03$ & $0.25\pm0.02$ & $0.26\pm0.02$ & $0.24\pm0.03$
\\
$\delta_{A_V}$ & $17^{+9}_{-12}$ & $349^{+10}_{-8}$ & $73\pm7$ & $355^{+13}_{-12}$ & $27^{+8}_{-9}$ & $68\pm8$
\\
$|A_P|$        & $0.22\pm0.03$ & $0.22\pm0.03$ & $0.19\pm0.03$ & $0.15\pm0.03$ & $0.14^{+0.03}_{-0.02}$
               & $0.16\pm0.03$
\\
$\delta_{A_P}$ & $342^{+12}_{-9}$ & $24^{+9}_{-11}$ & $108^{+9}_{-11}$ & $20^{+12}_{-27}$ & $13^{+45}_{-17}$
               & $98^{+11}_{-17}$
\\
$\chi^2_{\rm min}$ & $5.438$ & $5.603$ & $5.604$ & $7.345$ & $7.495$ & $7.956$
\\
Fit quality    & $0.1424$ & $0.1326$ & $0.1096$ & $0.062$ & $0.058$ & $0.047$
\\
\end{tabular}
\end{ruledtabular}
\label{fitting}}
\end{table}

\begin{table}[t]
\caption{Fit results using Eq.~(\ref{decaywidthC}) and $\phi=43.5^\circ$.  The amplitude sizes are quoted in units of $10^{-6}$ and the strong phases in units of degrees.
}
\label{tab:CFVPB}
\medskip
\footnotesize{
\begin{ruledtabular}
\begin{tabular}{l l c c c c c c c c c}
         &$|T_V|$                   &$|T_P|$            &$\delta_{T_P}$          &$|C_V|$                          &$\delta_{C_V}$          &$|C_P|$                  &$\delta_{C_P}$                &$|E_V|$                         &$\delta_{E_V}$     \\
          &$|E_P|$                            &$\delta_{E_P}$   &$|A_V|$                           &$\delta_{A_V}$             &$|A_P|$ &$\delta_{A_P}$ &$\chi^2_{\rm min}$ &fit quality\\
\hline
(S1)  &$2.18^{+0.06}_{-0.07}$ &$3.41\pm0.06$ &$69\pm3$ &$1.76\pm0.04$ &$278\pm3$ &$2.10\pm0.03$ & $201\pm1$     &$0.27\pm0.04$ & $260^{+50}_{-20}$  \\
         &$1.66^{+0.05}_{-0.06}$ & $108\pm3$    &$0.19\pm0.02$ &$17^{+9}_{-12}$          &$0.22\pm0.03$  &$342^{+12}_{-9}$          &$5.438$ &$0.1424$\\ 
            \hline
(S2)  &$2.18^{+0.06}_{-0.07}$ &$3.36\pm0.06$ &$286\pm3$ &$1.76\pm0.04$ &$76\pm3$ &$2.07\pm0.03$ &$201\pm1$  &$0.26\pm0.04$ &$69^{+46}_{-21}$  \\
          &$1.66^{+0.05}_{-0.06}$ &$108\pm3$    &$0.20\pm0.03$ &$349^{+10}_{-8}$          &$0.22\pm0.03$  &$24^{+9}_{-11}$          &$5.603$ &$0.1326$\\ 
            \hline
(S3) &$2.17\pm0.06$ &$3.51\pm0.06$ & $40^{+3}_{-4}$ &$1.74\pm0.04$ &$195^{+4}_{-3}$ &$2.04\pm0.03$ &$201\pm1$  &$0.40\pm0.06$ &$245^{+8}_{-9}$  \\
           &$1.66\pm0.05$ &$107\pm3$    &$0.22\pm0.03$ &$73\pm7$          &$0.19\pm0.03$  &$108^{+9}_{-11}$          &$5.604$ &$0.1096$\\ 
            \hline
(S4)  &$2.19^{+0.06}_{-0.07}$ &$3.48\pm0.06$ &$307^{+4}_{-3}$ &$1.75\pm0.04$ &$152^{+3}_{-4}$ &$2.14\pm0.03$ &$159\pm1$  &$0.33\pm0.05$ &$113^{+14}_{-11}$  \\
           &$1.66^{+0.05}_{-0.06}$ &$251\pm3$   &$0.25\pm0.02$ &$355^{+13}_{-12}$          &$0.15\pm0.03$  &$20^{+12}_{-27}$          &$7.345$ &$0.062$\\ 
            \hline
(S5)  &$2.18^{+0.06}_{-0.07}$ &$3.50\pm0.06$ &$79^{+3}_{-4}$ &$1.74\pm0.04$ &$235^{+4}_{-3}$ &$2.07\pm0.03$ &$159\pm1$     &$0.38\pm0.05$ &$282^{+8}_{-10}$\\
            &$1.66\pm0.05$ &$252\pm3$ &$0.26\pm0.02$ &$27^{+8}_{-9}$          &$0.14^{+0.03}_{-0.02}$  &$13^{+45}_{-17}$          &$7.495$ &$0.058$\\ 
             \hline
(S6)  &$2.18\pm0.06$ &$3.39\pm0.06$ &$12\pm3$ &$1.76\pm0.04$ &$221\pm3$ &$2.07\pm0.03$ &$201\pm1$    &$0.26\pm0.04$ &$224^{+22}_{-40}$  \\
            &$1.66^{+0.05}_{-0.06}$ &$108\pm3$ &$0.24\pm0.03$ &$68\pm8$          &$0.16\pm0.03$  &$98^{+11}_{-17}$          &$7.956$ &$0.047$\\ 
\end{tabular}
\end{ruledtabular}
\label{fitting}}
\end{table}

By performing a $\chi^2$ fit to the CF $D\to VP$ decays, we extract the magnitudes and strong phases of the topological amplitudes $T_V,C_V,E_V,A_V$ and $T_P,C_P,E_P,A_P$ from the measured partial widths through Eq.~(\ref{decaywidthA}) or (\ref{decaywidthC}) and find many possible solutions with local $\chi^2$ minima.  Here we take the convention that all strong phases are defined relative to the $T_V$ amplitude.  In 2016 we have performed a detailed analysis and obtained some best $\chi^2$ fit solutions (A) and (S) through Eqs.~(\ref{decaywidthA}) and (\ref{decaywidthC}), respectively~\cite{Cheng:2016}. It turns out that solutions (S) give a better description for SCS decays such as $D^0\to \pi^+\rho^-,\pi^0\rho^0$ and $D^+\to\pi^+\rho^0$, possibly because the additional SU(3)-breaking factor in phase space has been taken care of, as mentioned above. Hence, we will confine ourselves to using Eq.~\eqref{decaywidthC} and thus solutions (S) in this work.

The six best $\chi^2$-fit solutions (S1)--(S6), with $\chi^2_{\rm min} < 10$, are listed in Table~\ref{tab:CFVPB}, where we have chosen the convention such that the central values of strong phases to fall between 0 and 360 degrees, while noting again that a simultaneous sign flip of all strong phases is equally viable. The flavor amplitudes of all these solutions respect the hierarchy pattern, $|T_P|>|T_V|\gsim|C_{P}|>|C_{V}|\gsim|E_P|>|E_V|\gsim|A_{P,V}|$.  As stressed in~\cite{Cheng:2016}, the decay $D_s^+\to \rho^0\pi^+$ plays an essential role in the determination of the annihilation amplitudes $A_{V,P}$. Its large error in the branching fraction reflects in the large uncertainties in the magnitudes and strong phases of $A_{V,P}$, which will be improved once we have a better measurement of $D_s^+\to \rho^0\pi^+$.

While the size of each topological amplitude is similar across all solutions, the strong phases vary among the solutions except for those of $C_P$ and $E_P$. We find ($\delta _{C_P}$, $\delta_{E_P}$) to be either
(201$^\circ$, 108$^\circ$) or (159$^\circ$, 252$^\circ$).  A close inspection tells us that Solutions (S1) and (S4) are close to each other in the sense that the corresponding amplitudes are similar in size, except for $|A_V|$ and $|A_P|$, and the corresponding strong phases add up to roughly $360^\circ$.  So are Solutions (S2) and (S5).

\begin{table}[tp!]
\caption{Flavor amplitude decompositions, experimental branching fractions, and predicted branching fractions for the Cabibbo-favored $D \to VP$ decays.  Here $s_\phi\equiv\sin\!\phi$, $c_\phi\equiv\cos\!\phi$ and $\lambda_{sd}\equiv V_{cs}^*V_{ud}$.  The columns of ${\cal B}_{\rm theory}(S3)$ and ${\cal B}_{\rm theory}(S6)$ are predictions based on Solutions (S3) and (S6) shown in Table~\ref{tab:CFVPB}, respectively.  All branching fractions are quoted in units of \%.  }
\vspace{6pt}
\begin{ruledtabular}
\begin{tabular}{l l l c c c c c c}
Meson & Mode & Representation
& ${\cal B}_{\rm exp}$ & ${\cal B}_{\rm theory}(S3)$  & ${\cal B}_{\rm theory}(S6)$
\\
\hline
$D^0$  & $K^{*-}\,\pi^+$          & $\lambda_{sd}(T_V + E_P)$
&$5.34\pm0.41$                    &$5.39\pm0.40$		&$5.35\pm0.40$         
\\
&$K^-\,\rho^+$                      & $\lambda_{sd}(T_P + E_V)$
&$11.3 \pm 0.7$                    &$11.4\pm0.6$		&$11.7\pm0.8$         
\\
& $\ol{K}^{*0}\,\pi^0$             &$\frac{1}{\sqrt{2}}\lambda_{sd}(C_P - E_P)$
&$3.74\pm0.27$                     &$3.67\pm0.21$		&$3.69\pm0.21$         
\\
& $\ol{K}^0\,\rho^0$              & $\frac{1}{\sqrt{2}}\lambda_{sd}(C_V - E_V)$
&$1.26^{+0.12}_{-0.16}$           &$1.30\pm0.12$		&$1.35\pm0.13$         
\\
& $\ol{K}^{*0}\,\eta$               & $\lambda_{sd}\left[ {1\over \sqrt{2}}(C_P + E_P)c_\phi - E_Vs_\phi\, \right]$
&$1.02\pm0.30$                      &$0.92\pm0.08$		&$0.86\pm0.12$         
\\
& $\ol{K}^{*0}\,\eta\,'$            &$-\lambda_{sd}\left[ {1\over \sqrt{2}}(C_P + E_P)s_\phi + E_Vc_\phi\, \right]$
&$<0.10$                            &$0.0048\pm0.0004$		&$0.0052\pm0.0007$	
\\
& $\ol{K}^0\,\omega$             & $-\frac{1}{\sqrt{2}}\lambda_{sd}(C_V + E_V)$
&$2.22 \pm 0.12$                 &$2.23\pm0.16$		&$2.17\pm0.16$        
\\
& $\ol{K}^0\,\phi$                   &$-\lambda_{sd}E_P$
&$0.830\pm0.061$  			& $0.835\pm0.054$		&$0.838\pm0.054$      
\\
\hline
$D^+$
& $\ol{K}^{*0}\,\pi^+$  & $\lambda_{sd}(T_V + C_P)$
&$1.57\pm0.13$                   &$1.59\pm0.15$		&$1.58\pm0.15$  
\\
&$\ol{K}^0\,\rho^+$               & $\lambda_{sd}(T_P + C_V)$
&$12.3^{+1.2}_{-0.7}$             &$12.5\pm1.5$		&$12.3\pm1.5$ 	
\\
\hline
$D_s^+$
& $\ol{K}^{*0}\,K^+$& $\lambda_{sd}(C_P + A_V)$
&$3.92\pm0.14$                   &$3.94\pm0.18$		&$3.94\pm0.18$ 	
\\
&$\ol{K}^0\,K^{*+}$              & $\lambda_{sd}(C_V + A_P)$
&$5.4 \pm 1.2$                    &$3.39\pm0.21$		&$3.10\pm0.21$ 	
\\
&$\rho^+\,\pi^0$                   & $\frac{1}{\sqrt{2}}\lambda_{sd}(A_P - A_V)$
&---                                  &$0.024\pm0.014$		&$0.025\pm0.016$ 	
\\
&$\rho^+\,\eta$                    & $\lambda_{sd}\left[ {1\over\sqrt{2}}( A_P + A_V)c_\phi-T_Ps_\phi \right]$
&$8.9 \pm 0.8$                     &$9.02\pm0.37$		&$8.86\pm0.38$ 	
\\
&$\rho^+\,\eta\,'$                 & $\lambda_{sd} \left[{1\over\sqrt{2}}( A_P + A_V)s_\phi+T_Pc_\phi \right]$
&$5.8\pm1.5$   						&$3.25\pm0.12$		&$2.92\pm0.11$ 	
\\
& $\pi^+\,\rho^0$                 & $\frac{1}{\sqrt{2}}\lambda_{sd}(A_V - A_P)$
 &$0.020\pm0.012$               &$0.023\pm0.014$		&$0.024\pm0.016$ 	
\\
& $\pi^+\,\omega$                &$\frac{1}{\sqrt{2}}\lambda_{sd}(A_V + A_P)$
&$0.19\pm 0.03$\footnotemark[1]                 &$0.19\pm0.04$		&$0.19\pm0.04$		
\\
& $\pi^+\,\phi$                      &$\lambda_{sd}T_V$
&$4.5 \pm 0.4$                     &$4.45\pm0.24$		&$4.49\pm0.25$ 	
\\
\end{tabular}
\label{VPCF}
\footnotetext[1]{New measurement from BESIII~\cite{BESIII:DsKomega} has been taken into account in the world average.}
\end{ruledtabular}
\end{table}

Although solutions in set (S) generally fit the Cabibbo-favored modes well (see Table~\ref{VPCF} for results based on Solutions (S3) and (S6)), there are two exceptions, namely, $D_s^+\to \overline{K}^0K^{*+}$ and $\rho^+\eta'$, where the predictions are smaller than the experimental results.
The first mode was measured three decades ago with a relatively large uncertainty~\cite{CLEO:DsKKst}, and the experimental result was likely to be overestimated.  The second mode has a decay amplitude respecting a sum rule~\cite{Cheng:2016}:
\be
{\cal M}(D_s^+\to \pi^+\omega)=\cos\phi{\cal M}(D_s^+\to\rho^+\eta)+\sin\phi{\cal M}(D_s^+\to\rho^+\eta').
\en
Assuming this relation, the current data of $\B(D_s^+\to\pi^+\omega)$ and $\B(D_s^+\to\rho^+\eta)$ give the bounds
$1.6\% < \B(D_s^+\to\rho^+\eta') < 3.9\%$ at $1\sigma$ level, significantly lower than the current central value.  A better determination of these branching fractions will be very helpful in settling the issues.

Various (S) solutions lead to very different predictions for some of the SCS decays.
Especially, the $D^0\to\pi^0\omega$ and $D^+\to\pi^+\omega$ decays are very useful in discriminating among different solutions. We first consider the $\pi^0\rho^0,\pi^0\omega$ and $\eta\omega$ modes. Their topological amplitudes are given by
\be
{\cal M}(D^0\to \pi^0\omega) &=& {1\over 2}\lambda_d(C_V-C_P+E_P+E_V), \non \\
{\cal M}(D^0\to \pi^0\rho^0) &=& {1\over 2}\lambda_d(C_V+C_P-E_P-E_V),  \\
{\cal M}(D^0\to \eta\omega) &=& {1\over 2}\lambda_d(C_V+C_P+E_P+E_V)\cos\phi-{1\over\sqrt{2}}\lambda_s C_V\sin\phi. \non
\en
Since the magnitude of $C_V$ is comparable to that of $C_P$,  the smallness of $\B(D^0\to\pi^0\omega)$, the sizable $\B(D^0\to\eta\omega)$ and the large  $\B(D^0\to\pi^0\rho^0)$ imply that the strong phases of $C_V$ and $C_P$ should be close to each other. An inspection of Table~\ref{tab:CFVPB} indicates that the phase difference between $C_V$ and $C_P$ is large for Solutions (S1), (S2) and (S5). It turns out that (S2) and (S5) are definitely ruled out as they predict too large $\B(D^0\to\pi^0\omega)$, with the central values of $4.65$ and $3.91$ (in units of $10^{-3}$), respectively, while the measured value is $0.117\pm0.035$ (see Table~\ref{tab:CSPV}). Solution (S1) gives a relatively better prediction of $\B(D^0\to\pi^0\omega)=0.62\pm0.13$ among the three solutions.

We next turn to the $\pi^+\rho^0$ and $\pi^+\omega$ modes. Neglecting the penguin contributions, their topological amplitudes read (see Table~\ref{tab:CSPV})
\be \label{eq:D+pirho}
{\cal M}(D^+\to \pi^+\rho^0) &=& {1\over \sqrt{2}}\lambda_d (T_V+C_P-A_P+A_V), \non \\
{\cal M}(D^+\to \pi^+\omega) &=& {1\over \sqrt{2}}\lambda_d (T_V+C_P+A_P+A_V).
\en
It is well known that the CF decays $D_s^+\to \pi^+\rho^0$ and $\pi^+\omega$ can only proceed through the $W$-annihilation topology
\be
{\cal M}(D_s^+\to \pi^+\rho^0) &=& {1\over \sqrt{2}}V_{cs}^*V_{ud}(A_V-A_P), \non \\
{\cal M}(D_s^+\to \pi^+\omega) &=& {1\over \sqrt{2}}V_{cs}^*V_{ud} (A_V+A_P).
\en
The extremely small branching fraction of $D_s^+\to\pi^+\rho^0$ compared to $D_s^+\to \pi^+\omega$ (see Table~\ref{tab:CFVPB}) implies that $A_V$ and $A_P$ should be comparable in magnitude and roughly parallel to each other with a phase difference not more than $30^\circ$. At a first glance, it is tempting to argue from Eq.~(\ref{eq:D+pirho}) that $D^+\to\pi^+\omega$ should have a rate larger than $D^+\to\pi^+\rho^0$. Experimentally, it is the other way around~\cite{PDG}:
\be  \label{eq:BFD+pirho}
\B(D^+\to\pi^+\rho^0)=(0.83\pm0.15)\times 10^{-3}, \qquad
\B(D^+\to \pi^+\omega)=(0.28\pm0.06)\times 10^{-3}.
\en
Since $C_P$ is comparable to $T_V$ in magnitude, there is a large cancellation between $T_V$ and $C_P$. As a consequence, the rates of $\pi^+\rho^0$ and $\pi^+\omega$ become sensitive to the strong phases of the small annihilation amplitudes $A_V$ and $A_P$. It turns out that $A_V$ should be in the 4th quadrant while $A_P$ in the third quadrant in order to satisfy the experimental constraints from Eq.~(\ref{eq:BFD+pirho}). We find only Solutions (S3) and (S6) in line with this requirement (see Table~\ref{tab:CFVPB}) and yielding predictions in agreement with experiment for $\pi^+\rho^0$ and $\pi^+\omega$ (see Table~\ref{tab:CSPV}). For Solutions (S1), (S2), (S4) and (S5),  the branching fractions of $D^+\to \pi^+\rho^0$ and $D^+\to \pi^+\omega$ (in units of $10^{-3}$) are found to have the central values (0.45, 1.06), (0.87, 0.98), (0.67, 1.05), (0.96, 1.76), respectively. All these solutions imply that the latter is larger than the former in rates, in contradiction with experiment.

Finally, we comment on two of the $D_s^+$ decay modes: $K^+\rho^0$ and $K^+\omega$. From Table~\ref{tab:CSPV}, we see that
\be \label{eq:DsKrho}
{\cal M}(D_s^+\to K^+\rho^0) = {1\over \sqrt{2}}(\lambda_d C_P-\lambda_s A_P) , \qquad
{\cal M}(D_s^+\to K^+\omega) = {1\over \sqrt{2}}(\lambda_d C_P+\lambda_s A_P).
\en
Since $|C_P|\gg|A_P|$, it is expected that the two modes have similar branching fractions of order $2\times 10^{-3}$. However, the recent BESIII experiment yields
$\B(D_s^+\to K^+\omega)=(0.87\pm0.25)\times 10^{-3}$~\cite{BESIII:DsKomega}. The $\rho-\omega$ mixing effect to be mentioned below in Eq.~(\ref{eq:rhoomegamix}) in principle can push up (down) the rate of $K^+\rho^0$ ($K^+\omega$). For the mixing angle $\epsilon=-0.12$ (see Eq.~(\ref{eq:rhoomegamix}) and note a sign difference from~\cite{Qin}), we find $\B(D_s^+\to K^+\rho^0)=(2.63\pm0.11)\times 10^{-3}$ and $\B(D_s^+\to K^+\omega)=(1.65\pm0.09)\times 10^{-3}$. The former is now in better agreement with experiment, but the latter is still too large compared to the data. The $\omega-\phi$ mixing also does not help much. Moreover, in our framework we do not need $\rho-\omega$ mixing to explain the smallness of $D^0\to\pi^0\omega$ and $D^+\to \pi^+\omega$. Therefore, the issue with $D_s^+\to K^+\omega$ remains to be resolved.

\renewcommand{\arraystretch}{1.00}
\begin{table}[tp!]
\caption{Same as Table~\ref{VPCF}, but for the singly Cabibbo-suppressed decay modes.
 All branching fractions are quoted in units of $10^{-3}$.
  \label{tab:CSPV}}
\medskip
  \scriptsize{
\begin{ruledtabular}
\begin{tabular}{l l l c c  c c }
 & Mode & Representation & ${\cal B}_{\rm exp}$ & ${\cal B}_{\rm theo}$(S3)
  & ${\cal B}_{\rm theo}$(S6) \\
\hline
$D^0$
  & $\pi^+ \rho^-$ & $\lambda_d (T_V+E_P)
  + \lambda_p( P_V^p+ \PAP + \PEP )$
     & $5.15 \pm 0.25$  & $4.72\pm0.35$ & $4.68 \pm 0.35$ \\
  & $\pi^- \rho^+$ & $\lambda_d  (T_P+E_V)
  + \lambda_p( P_P^p+ \PAV + \PEV )$
     & $10.1 \pm 0.4$ & $8.81\pm0.46$ & $9.14 \pm 0.60$ \\
  & $\pi^0 \rho^0$ & $\frac12 \lambda_d (-C_P-C_V+E_P+E_V)$
     & $3.86 \pm 0.23$ & $3.18\pm0.19$ & $3.92 \pm 0.20$ \\
  && \quad $ +\lambda_p(P_P^p + P_V^p + \PAP + \PAV + \PEP +\PEV )$ \\
  & $K^+ K^{*-}$ & $\lambda_s (T_V+E_P) + \lambda_p(P_V^p + \PEP + \PAP)$
     & $1.65 \pm 0.11$  & $1.81 \pm 0.14$ & $1.79 \pm 0.13$ \\
  & $K^- K^{*+}$ & $\lambda_s (T_P+E_V) +\lambda_p( P_P^p + \PEV + \PAV)$
     & $4.56 \pm 0.21$  & $3.35\pm0.17$ & $3.44 \pm 0.23$ \\
  & $K^0 \ol{K}^{*0}$ & $\lambda_d E_V+ \lambda_s E_P +  \lambda_p(\PAP + \PAV)$
     & $0.246\pm0.048$  & $1.27\pm0.10$ & $1.04 \pm 0.14$   \\
  & $\ol{K}^0 K^{*0}$ &  $\lambda_d E_P+ \lambda_s E_V + \lambda_p( \PAP + \PAV)$
     & $0.336\pm0.063$  & $1.27\pm0.10$ & $1.04 \pm 0.14$ \\
  & $\pi^0 \omega$ & $\frac{1}{2}\lambda_d (-C_V+C_P-E_P-E_V)
  + \lambda_p(P_P^p + P_V^p + \PEP + \PEV)$
     & $0.117\pm0.035$ & $0.53\pm0.09$  & $0.22 \pm 0.06$\\
  & $\pi^0 \phi$ & $\frac{1}{\sqrt{2}} \lambda_s C_P$
     & $1.20 \pm 0.04$\footnotemark[1]  & $0.64 \pm 0.02$ & $0.65 \pm 0.02$ \\
  & $\eta \omega$ & $\frac12 [\lambda_d (C_V+C_P+E_V+E_P)\cos\phi
   - \lambda_s C_V\sin\phi$
     & $1.98\pm0.18$  & $2.96\pm0.13$ & $2.56 \pm 0.14$  \\
  & & $\quad +\lambda_p(P_P^p + P_V^p +\PEP + \PEV + \PAP + \PAV) \cos\phi]$ &&& \\
  & $\eta\,' \omega$ & $\frac12 [\lambda_d(C_V+C_P+E_V+E_P)\sin\phi +\lambda_s C_V\cos\phi$
     &--- & $0.03\pm0.00$ &  $0.05 \pm 0.01$  \\
   & & $\quad +\lambda_p(P_P^p + P_V^p +\PEP + \PEV + \PAP + \PAV) \sin\phi]$ &&& \\
  & $\eta \phi$ & $\lambda_s [{1\over\sqrt{2}}C_P\cos\phi - (E_V+E_P)\sin\phi ]
  + \lambda_p(\PAP + \PAV) \sin\phi $
     & $0.167\pm0.034$\footnotemark[1] & $0.24\pm0.02$ & $0.29 \pm 0.03$ \\
  & $\eta \rho^0$ & ${1\over 2}[\lambda_d (C_V-C_P-E_V-E_P)\cos\phi
  - \lambda_s\sqrt{2}C_V\sin\phi$
     & --- & $0.31\pm0.05$ &  $0.84 \pm 0.10$   \\
    & & $\quad + \lambda_p(P_P^p + P_V^p + \PEP + \PEV) \cos\phi]$ &&& \\
  & $\eta\,' \rho^0$ & ${1\over 2}[\lambda_d (C_V-C_P-E_V-E_P)\sin\phi
  + \lambda_s\sqrt{2}C_V\cos\phi$
     & ---  & $0.11 \pm 0.01$ & $0.10 \pm 0.01$ \\
      & & $\quad + \lambda_p(P_P^p + P_V^p + \PEP + \PEV) \sin\phi]$ &&& \\
\hline
$D^+$
  & $\pi^+ \rho^0$ & $\frac{1}{\sqrt{2}}[\lambda_d(T_V+C_P-A_P+A_V)
  + \lambda_p(P_V^p - P_P^p + \PEP - \PEV)]$
     & $0.83 \pm 0.15$ & $0.70\pm0.10$ & $0.61\pm0.10$  \\
  & $\pi^0 \rho^+$ & $\frac{1}{\sqrt{2}}[\lambda_d (T_P+C_V+A_P-A_V)
  + \lambda_p(P_P^p - P_V^p + \PEV - \PEP)]$
     &---  & $4.43\pm0.61$ & $4.53\pm0.64$\\
  & $\pi^+ \omega$ & $\frac{1}{\sqrt{2}}[\lambda_d (T_V+C_P+A_P+A_V)
  + \lambda_p(P_P^p + P_V^p + \PEP + \PEV)]$
     &$0.28\pm0.06$ & $0.22\pm0.06$  & $0.26\pm0.07$ \\
  & $\pi^+ \phi$ & $\lambda_s C_P$
     &$5.68\pm0.11$\footnotemark[1]  & $3.27\pm0.11$  & $3.35 \pm 0.11$  \\
  & $\eta \rho^+$ & ${1\over\sqrt{2}}[\lambda_d (T_P+C_V+A_V+A_P)\cos\phi
   - \lambda_s\sqrt{2}C_V\sin\phi$
     & --- & $1.53\pm0.49$ &  $1.02\pm0.34$ \\
  &&$\quad + \lambda_p(P_P^p + P_V^p + \PEP + \PEV) \cos\phi]$ &&& \\
  & $\eta\,' \rho^+$ & ${1\over\sqrt{2}}[\lambda_d (T_P+C_V+A_V+A_P)\sin\phi
  + \lambda_s\sqrt{2}C_V\cos\phi$
     & --- & $1.16\pm0.11$ &  $1.03\pm0.11$ \\
 &&$\quad + \lambda_p(P_P^p + P_V^p + \PEP + \PEV) \sin\phi]$&&& \\
  & $K^+ \ol{K}^{*0}$ & $\lambda_d A_V + \lambda_s T_V +  \lambda_p(P^p_V + \PEP)$
     & $3.83^{+0.14}_{-0.21}$ & $3.87\pm0.23$ & $3.82\pm0.25$  \\
  & $\ol{K}^0 K^{*+}$ & $\lambda_d A_P + \lambda_s T_P + \lambda_p( P_P^p + \PEV)$
     & $34 \pm 16$ & $10.20\pm0.40$ & $9.80\pm0.41$  \\
\hline
$D_s^+$
  & $\pi^+ K^{*0}$ & $\lambda_d T_V + \lambda_s A_V + \lambda_p(P_V^p + \PEP)$
     & $2.13 \pm 0.36$  & $3.69\pm0.23$ & $3.65\pm0.24$ \\
  & $\pi^0 K^{*+}$ & $\frac{1}{\sqrt{2}}[\lambda_d C_V - \lambda_s A_V - \lambda_p(P_V^p + \PEP) ]$
     & --- & $1.12\pm0.07$ & $1.02\pm0.07$  \\
  & $K^+ \rho^0$ & $\frac{1}{\sqrt{2}}[\lambda_d C_P - \lambda_s A_P - \lambda_p(P_P^p + \PEV) ]$
     & $2.5 \pm 0.4$ & $2.10\pm0.10$ & $2.10\pm0.10$  \\
  & $K^0 \rho^+$ & $\lambda_d T_P + \lambda_s A_P + \lambda_p(P_P^p + \PEV)$
     & --- & $11.80\pm0.47$ &  $11.47\pm0.48$  \\
  & $\eta K^{*+}$ & ${1\over\sqrt{2}}\big\{  [\lambda_d C_V +  \lambda_s A_V
  + \lambda_p(P_V^p + \PEP )]\cos\phi$
     & --- & $0.60\pm0.21$ &  $0.64\pm0.20$ \\
  &&$\quad - [ \lambda_s(T_P+C_V+A_P) + \lambda_p(P_P^p + \PEV) ]\sin\phi\big\}$&&& \\
  & $\eta\,' K^{*+}$
  & ${1\over\sqrt{2}}\big\{ [\lambda_d C_V + \lambda_s A_V + \lambda_p(P_V^p + \PEP)] \sin\phi$
     & --- & $0.38\pm0.02$ &  $0.33\pm0.02$ \\
  &&$\qquad - [ \lambda_s(T_P+C_V+A_P) + \lambda_p(P_P^p + \PEV)]\cos\phi \big\}$&  && \\
  & $K^+ \omega$ & $\frac{1}{\sqrt{2}}\left[ \lambda_d C_P + \lambda_s A_P
  + \lambda_p(P_P^p + \PEV) \right]$
     & $0.87\pm0.25$\footnotemark[2]  & $2.02\pm0.09$ & $2.12\pm0.10$ \\
  & $K^+ \phi$ & $\lambda_s(T_V+C_P+A_V) + \lambda_p(P_V^p + \PEP)$
     & $0.182\pm0.041$  & $0.13\pm0.02$ & $0.12\pm0.02$  \\
\end{tabular}
\footnotetext[1]{New measurements from BESIII~\cite{BESIII:DtophiP} have been taken into account in the world average.}
\footnotetext[2]{Data from BESIII~\cite{BESIII:DsKomega}.}
\end{ruledtabular} }
\end{table}
%

\subsection{Flavor SU(3) symmetry breaking}

As noted in passing, a most noticeable example of SU(3) breaking in the $P\!P$ sector lies in the decays $D^0\to K^+K^-$ and $D^0\to \pi^+\pi^-$. Experimentally, the rate of the former is larger than that of the latter by a factor of $2.8$\,. More precisely,  $|T+E|_{K\!K}/|T+E|_{\pi\pi}\approx 1.80$, implying a large SU(3) breaking effect in the amplitude of $T+E$.  However, it is the other way around for the counterparts in the $V\!P$ sector where we have $\Gamma(K^+K^{*-})<\Gamma(\pi^+\rho^-)$ and $\Gamma(K^-K^{*+})<\Gamma(\pi^-\rho^+)$. Since the available phase space is proportional to $p_c^3/m_V^2$ in the convention of Eq.~(\ref{decaywidthC}), this explains why $\Gamma(D^0\to KK^*)<\Gamma(D^0\to \pi\rho)$ owing to the fact that $p_c(\pi\rho)=764$~MeV and $p_c(KK^*)=608$~MeV. From the measured branching fractions, we find by ignoring the penguin amplitudes that
\be \label{eq:TEinVP}
{|T_V+E_P|_{\pi^+\rho^-}\over |T_V+E_P|_{K^+K^{*-}} }=1.08\,, \qquad {|T_P+E_V|_{\pi^-\rho^+}\over |T_V+E_P|_{K^-K^{*+}} }=0.91\,.
\en
This implies that SU(3) breaking in the amplitudes
of $T_V+E_P$ and $T_P+E_V$ is small, contrary to the $P\!P$ case.

In Table~\ref{tab:CSPV}, we show the calculated branching fractions of SCS $D\to VP$ decays using Solutions (S3) and (S6). It is clear that Solution (S6) is slightly better, though the predicted $K^0\ol K^{*0}$ and $\ol K^0 K^{*0}$ branching fractions are too large compared to the data in both solutions. SU(3) breaking effects in the color-allowed and color-suppressed amplitudes can be estimated provided they are factorizable:
\be
T_V &=& \frac{G_F}{\sqrt2}a_1(\ol K^*\pi)2f_\pi m_{K^*}A_0^{DK^*}(m_\pi^2) ~, \non \\
C_P &=& \frac{G_F}{\sqrt2}a_2(\ol K^*\pi)2f_{K^*} m_{K^*}F_1^{D\pi}(m_{K^*}^2) ~,  \non \\
T_P &=& \frac{G_F}{\sqrt2}a_1(\ol K\rho)2f_\rho m_\rho F_1^{DK}(m_\rho^2) ~,\\
C_V &=& \frac{G_F}{\sqrt2}a_2(\ol K\rho)2f_K m_\rho A_0^{D\rho}(m_K^2) ~. \non
\label{factorizationS}
\en
Hence,
\be
{T_V^{\pi\rho}\over T_V}={a_1(\rho\pi)\over a_1(\bar K^*\pi)}{m_\rho\over m_{K^*}} {A_0^{D\rho}(m_\pi^2)\over A_0^{DK^*}(m_\pi^2)},  \qquad {T_P^{\pi\rho}\over T_P}={a_1(\rho\pi)\over a_1(\bar K\rho)} {F_1^{D\pi}(m_\rho^2)\over F_1^{DK}(m_\rho^2)}.
\en
Assuming that $a_1(\rho\pi)$ is similar to $a_1(\bar K^*\pi)$ and $a_1(\bar K\rho)$, we find $T_V(\pi^+\rho^-) \simeq 0.82\,T_V$, $T_P(\pi^-\rho^+) \simeq 0.92\,T_P$,
$T_V(K^+K^{*-}) \simeq 1.29\,T_V$ and $T_P(K^-K^{*-}) \simeq 1.28\,T_P$.  Similar relations can be derived for the $C_V$ and $C_P$ amplitudes as well. These lead to two difficulties: (i) The sizable SU(3) breaking in the ratios $|T_V|_{\pi^+\rho^-}/ |T_V|_{K^+K^{*-}} \simeq 0.64$ and $|T_P|_{\pi^-\rho^+}/ |T_V|_{K^-K^{*+}} \simeq 0.72$ are not consistent with Eq.~(\ref{eq:TEinVP}), and (ii)
the branching fractions of $D^0\to \pi^+\rho^-$ and $D^0\to \pi^-\rho^+$ will become smaller, while $\B(D^0\to K^+K^{*-})$ and $\B(D^0\to K^-K^{*+})$ become larger.  Hence, the discrepancy becomes even worse. In other words, the consideration of SU(3) breaking in the tree amplitudes $T_{V,P}$ and $C_{V,P}$ alone will render even larger deviations from the data in both Solutions (S3) and (S6).

A way out is to consider SU(3) breaking in the $W$-exchange amplitudes. Indeed, the too large rates predicted for $K^0\ol K^{*0}$ and $\ol K^0 K^{*0}$ modes call for SU(3) breaking in the $W$-exchange amplitudes as both modes proceed through $E_P$ and $E_V$. In the $PP$ sector, we need SU(3) breaking in $W$-exchange in order to induce $D^0\to K_SK_S$. Here we need SU(3) breaking  again for a different reason, otherwise, the calculated $D^0\to K^0\ol K^{*0}$ and $\ol K^0 K^{*0}$ will be too large in rates.
Since $|E_P|\gg|E_V|$,
it is natural to expect that $|E_P|$ ($|E_V|$) has to be reduced (increased) after SU(3) breaking in order to accommodate the data. Writing
\be
{\cal M}(D^0\to \pi^+\rho^-)=\lambda_d(T_V+E_P^d),  && {\cal M}(D^0\to \pi^-\rho^+)=\lambda_d(T_P+E_V^d), \non \\
{\cal M}(D^0\to K^+K^{*-})=\lambda_s(T_V+E_P^s), && {\cal M}(D^0\to K^-K^{*+})=\lambda_s(T_P+E_V^s), \non \\
{\cal M}(D^0\to K^0\ol K^{*0})=\lambda_sE_P^s+\lambda_d E_V^d, && {\cal M}(D^0\to \ol K^0 K^{*0})=\lambda_sE_V^s+\lambda_d E_P^d, \nonumber
\en
and
\be
{\cal M}(D^0\to \pi^0\rho^0) &=& {1\over 2}\lambda_d(C_P+C_V-E_P^d-E_V^d), \non \\
{\cal M}(D^0\to \pi^0\omega) &=& {1\over 2}\lambda_d(C_V-C_P+E_P^d+E_V^d),
\en
with
\be
E_V^d=e_V^d e^{i\delta e_V^d}E_V, \quad E_V^s=e_V^s e^{i\delta e_V^s}E_V, \quad E_P^d=e_P^d e^{i\delta e_P^d}E_P, \quad E_P^s=e_P^se^{i\delta e_P^s} E_P,
\en
we are able to determine the eight unknown parameters $e_V^d,e_P^d,e_V^s,e_P^s$ and $\delta e_V^d,\delta e_P^d,\delta e_V^s,\delta e_P^s$ from the branching fractions of these eight modes. In the SU(3) limit, $e_{V,P}^{d,s}=1$ and $\delta e_{V,P}^{d,s}=0$.
Note that among all the best-fit solutions (S), only (S3) and (S6) give exact solutions
for the parameters $e_{V,P}^{d,s}$ and the phases $\delta e_{V,P}^{d,s}$ ({\it i.e.}, $\chi^2=0$ in a fit to the eight SCS modes).
There are six solutions for Solution (S6), listed in Table~\ref{tab:EPEV}.  All these schemes are equally good in explaining the first eight SCS modes in Table~\ref{tab:CSPV}, whereas Scheme (iv) yields smallest SU(3) symmetry violation in $e_{V,P}^{d,s}$; namely, the deviations of them from unity are less than 50\%.

\begin{table}[t]
\caption{Solutions of the parameters $e_{V,P}^{d,s}$ and the phases $\delta e_{V,P}^{d,s}$ describing SU(3) breaking effects in the $W$-exchange amplitudes for Solution (S6).
  \label{tab:EPEV}}
  \medskip
\begin{ruledtabular}
\begin{tabular}{l c  c c c c c c c}
 &  $e_V^d$ & $\delta e_V^d$ & $e_P^d$ & $\delta e_P^d$ & $e_V^s$ & $\delta e_V^s$ & $e_P^s$ & $\delta e_P^s$  \\
\hline
(i) & 1.50 & 241 & 0.18 & 290 & 3.44 & 69 & 0.29 & 159 \\
(ii) & 1.50 & 241 & 0.18 & 290 & 3.44 & 69 & 0.76 & 358 \\
(iii) & 1.12 & 55 & 0.51 & 336 & 6.67 & 243 & 6.35 & 347 \\
(iv) & 1.12 & 55 & 0.51 & 336 & 1.30 & 111 & 0.68 & 149 \\
(v) & 2.09 & 53 & 1.03 & 356 & 2.90 & 222 & 0.19 & 341 \\
(vi)  & 2.09 & 53 & 1.03 & 356 & 2.90 & 222 & 0.81 & 146 \\
\end{tabular}
\end{ruledtabular}
\end{table}

\begin{table}[tp!]
\caption{Branching fractions (in units of $10^{-3}$) of $D\to V\!P$ decays. The predictions  made in the (S6) scheme have taken into account SU(3) breaking effects under solution (iv) (see Table~\ref{tab:EPEV}).
For QCD-penguin exchanges $\PE_V$ and $\PE_P$, we assume that they are similar to the topological $E_V$ and $E_P$ amplitude, resepctively [see Eq.~(\ref{eq:PEVP})]. The results
from~\cite{Qin} in the factorization-assisted topological approach without and with the $\rho-\omega$ mixing (denoted by FAT and FAT[mix], respectively) are listed for comparison.
  \label{tab:BFVP}  }
\medskip
\begin{tabular}{l l c c c c }\hline\hline
 & Mode~~~ & ${\cal B}$(This work)~~ &  ~${\cal B}$(FAT)~ & ~${\cal B}$(FAT[mix])~ & ${\cal B}_{\rm exp}$ \\
\hline
$D^0$
  & $\pi^+ \rho^-$ & $5.12\pm0.29$ & 4.74 & 4.66 & $5.15 \pm 0.25$ \\
  & $\pi^- \rho^+$ & $10.21\pm0.91$  & 10.2 & 10.0 & $10.1 \pm 0.4$ \\
  & $\pi^0 \rho^0$ & $3.90\pm0.26$ & 3.55 & 3.83 & $3.86 \pm 0.23$  \\
  & $K^+ K^{*-}$ & $1.68\pm0.11$  & 1.72 & 1.73 & $1.65 \pm 0.11$ \\
  & $K^- K^{*+}$ & $4.43\pm0.31$ & 4.37 & 4.37 & $4.56 \pm 0.21$  \\
  & $K^0 \ol{K}^{*0}$ & $0.27\pm0.06$  & 1.1 & 1.1 & ~$0.246\pm0.048$~  \\
  & $\ol{K}^0 K^{*0}$ & $0.32\pm0.09$ & 1.1 & 1.1 & $0.336\pm0.063$  \\
  & $\pi^0 \omega$ & $0.12\pm0.05$  & 0.85 &  0.18 & $0.117\pm0.035$ \\
  & $\pi^0 \phi$ & $1.22\pm0.04$   & 1.11 & 1.11 &  $1.20 \pm 0.04$  \\
  & $\eta \omega$ & $2.25\pm0.14$ & 2.4 & 2.0 & $1.98\pm0.18$ \\
  & $\eta\,' \omega$ & $0.01\pm0.00$  & 0.04 & 0.02 &--- \\
  & $\eta \phi$ & $0.16\pm0.02$ & 0.19 & 0.18 & $0.167\pm0.034$ \\
  & $\eta \rho^0$ & $0.59\pm0.07$ & 0.54 & 0.45 & ---  \\
  & $\eta\,' \rho^0$ & $0.06\pm0.01$ & 0.21 & 0.27 & ---  \\
\hline
$D^+$
  & $\pi^+ \rho^0$ & $0.61\pm0.10$ & 0.42 & 0.58 & $0.83 \pm 0.15$\\
  & $\pi^0 \rho^+$ & $4.53\pm0.64$ & 2.7 & 2.5 & --- \\
  & $\pi^+ \omega$ & $0.26\pm0.07$ & 0.95 & 0.80 & $0.28\pm0.06$   \\
  & $\pi^+ \phi$ & $6.29\pm0.20$    & 5.65 & 5.65 & $5.68\pm0.11$    \\
  & $\eta \rho^+$ & $1.02\pm0.34$   & 0.7 & 2.2 & ---  \\
  & $\eta\,' \rho^+$ & $1.03\pm0.11$  & 0.7 & 0.8 & ---  \\
  & $K^+ \ol{K}^{*0}$ & $3.82\pm0.25$  & 3.61 & 3.60 & $3.83^{+0.14}_{-0.21}$  \\
  & $\ol{K}^0 K^{*+}$ & $9.80\pm0.41$  & 11 & 11 & $34 \pm 16$   \\
\hline
$D_s^+$
  & $\pi^+ K^{*0}$ & $3.65\pm0.24$ & 2.52 & 2.35 & $2.13 \pm 0.36$ \\
  & $\pi^0 K^{*+}$ & $1.02\pm0.07$  & 0.8 &  1.0 & --- \\
  & $K^+ \rho^0$ & $2.10\pm0.10$  & 1.9 & 2.5 & $2.5 \pm 0.4$  \\
  & $K^0 \rho^+$ &  $11.47\pm0.48$ & 9.1 & 9.6 & ---  \\
  & $\eta K^{*+}$ & $0.64\pm0.20$   & 0.2 & 0.2 & ---  \\
  & $\eta\,' K^{*+}$ & $0.33\pm0.02$  & 0.2 & 0.2 & ---  \\
  & $K^+ \omega$ & $2.12\pm0.10$  & 0.6 & 0.07 & $0.87\pm0.25$  \\
  & $K^+ \phi$ & $0.12\pm0.02$ & 0.166 & 0.166 & $0.182\pm0.041$   \\
  \hline\hline
\end{tabular}
\end{table}
%

Branching fractions of SCS $D\to VP$ decays are predicted in Table~\ref{tab:BFVP} using the topological amplitudes given in Solution (S6). For SU(3) breaking effects in $W$-exchange amplitudes $E_V$ and $E_P$, we specifically choose solution (iv) for SU(3) breaking parameters given in Table~\ref{tab:EPEV}, though the results are very similar in other schemes. The decays $D^0\to \pi^0\phi$ and $D^+\to\pi^+\phi$ are special as they proceed only through the internal $W$-emission diagram $C_P$. Its SU(3) breaking can be estimated from Eq.~(\ref{factorizationS}) to be
\be
{C_P^{\pi\phi}\over C_P}={f_\phi\over f_{K^*}}{m_\phi\over m_{K^*}}{F_1^{D\pi}(m_\phi^2)\over
F_1^{D\pi}(m_{K^*}^2)}.
\en
For the $q^2$ dependence of the form factor we use
\be
F_1^{D\pi}(q^2)={ F_1^{D\pi}(0) \over \left[1-(q^2/ m_{D^*}^2)\right]\left[1-\alpha_1^{D\pi}(q^2/m_{D^*}^2)\right] },
\en
with $F_1^{D\pi}(0)=F_0^{D\pi}(0)=0.666$ and $\alpha_1^{D\pi}=0.24$, and find $C_P^{\pi\phi}=1.37C_P$. The resulting $\B(D^0\to\pi^0\phi)=(1.22\pm0.04)\times 10^{-3}$ and $\B(D^+\to\pi^+\phi)=(6.29\pm0.21)\times 10^{-3}$ are consistent with experiment, though the latter is slightly large in the central value.

\vskip 0.2cm
\noindent \underline{Comparison with the work of Qin {\it et al.}~\cite{Qin}}
\vskip 0.2cm
In Table~\ref{tab:BFVP}, we have compared our results of SCS $D\to V\!P$ branching fractions with that in the factorization-assisted topological approach~\cite{Qin} without and with the $\rho-\omega$ mixing, denoted by FAT and FAT[mix], respectively. The predicted $\B(D^0\to \pi^0\rho^0)=0.85$ (in units of $10^{-3}$) in FAT is far too large compared to the data of $0.117\pm0.035$. In order to resolve this discrepancy, Qin {\it et al.} considered the $\rho-\omega$ mixing defined by
\be \label{eq:rhoomegamix}
|\rho^0\ra= |\rho^0_I\ra-\epsilon|\omega_I\ra, \qquad
|\omega\ra= \epsilon|\rho^0_I\ra+|\omega_I\ra,
\en
where $|\rho^0_I\ra$ and $|\omega_I\ra$ denote the isospin eigenstates. Using the mixing angle $\epsilon=0.12$, the predicted branching fraction of $D^0\to \pi^0\omega$ ia reduced to 0.18, while $\B(D^0\to \pi^0\rho^0)$ is increased from 3.55 to 3.83. However, the calculated $\B(D^+\to \pi^+\omega)=0.80$ after taking into account of $\rho-\omega$ mixing is still too large compared to the experimental value of $0.28\pm0.06$. As for the $D_s^+\to K^+\omega$ mode, it appears that the predicted branching fraction of 0.6 before $\rho-\omega$ mixing agrees with the data of $0.87\pm0.25$~\cite{BESIII:DsKomega}, while the predicted value of 0.07 after the mixing effect is far too small. Therefore, irrespective of $\rho-\omega$ mixing, $D^0\to \pi^0\omega$ and $D_s^+\to K^+\omega$ cannot be explained simultaneously in the FAT or modified FAT approach.

\begin{table}[tp!]
\caption{Same as Table~\ref{tab:BFVP} except for the direct \CP asymmetries of $D\to VP$ decays in units of $10^{-3}$, where $a_{\rm dir}^{({\rm tree})}$ denotes \CP asymmetry arising from purely tree amplitudes. The superscript (t+p) denotes tree plus QCD-penguin amplitudes, (t+pa) for tree plus weak penguin-annihilation ($P\!E$ and $P\!A$) amplitudes and ``tot'' for the total amplitude.
  \label{tab:CPVP}  }
\medskip
\begin{ruledtabular}
\begin{tabular}{l l  c r r r c}
 & Mode
  &  $a_{\rm dir}^{({\rm tree})}$~ &  $a_{\rm dir}^{({\rm t+p})}$~~~ &  $a_{\rm dir}^{({\rm t+pa})}$~~~
     & $a_{\rm dir}^{({\rm tot})}$(This work) & $a_{\rm dir}^{({\rm tot})}$\cite{Qin} \\
\hline
$D^0$
  & $\pi^+ \rho^-$ & 0 & $0.01\pm0.00$ & $0.76\pm0.22$ & $0.77\pm0.22$ & $-0.03$ \\
  & $\pi^- \rho^+$ & 0 & $-0.09\pm0.01$ & $-0.05\pm0.04$ & $-0.14\pm0.04$ & $-0.01$ \\
  & $\pi^0 \rho^0$ & 0 & $-0.03\pm0.00$ & $0.40\pm0.15$ & $0.37\pm0.15$ & $-0.03$ \\
  & $K^+ K^{*-}$ & 0 & $-0.19\pm0.01$ & $-0.56\pm0.37$ & $-0.75\pm0.37$ & $-0.01$  \\
  & $K^- K^{*+}$ & 0 & $0.11\pm0.01$ & $0.05\pm0.04$ & $0.15\pm0.04$ & 0  \\
  & $K^0 \ol{K}^{*0}$ & $-0.15\pm0.21$ & $-0.15\pm0.21$ & $-0.15\pm0.21$ & $-0.15\pm0.21$ & $-0.7$ \\
  & $\ol{K}^0 K^{*0}$ &  $-0.34\pm0.16$ & $-0.34\pm0.16$ & $-0.34\pm0.16$ & $-0.34\pm0.16$ & $-0.7$ \\
  & $\pi^0 \omega$ & 0 & $0.18\pm0.04$ & $-2.31\pm0.96$ & $-2.14\pm0.95$ & 0.02 \\
  & $\pi^0 \phi$ & 0 & 0~~~~~~ & 0~~~~~~ & 0~~~~~~ & $-0.0002$ \\
  & $\eta \omega$ &  $-0.10\pm0.01$ & $-0.08\pm0.01$ & $-0.40\pm0.10$ & $-0.38\pm0.10$ & $-0.1$  \\
  & $\eta\,' \omega$ &   ~~$2.40\pm0.34$ & $1.91\pm0.25$ & $1.42\pm0.71$ & $0.96\pm0.66$ & 2.2\\
  & $\eta \phi$ & 0 & 0~~~~~~ & 0~~~~~~ & 0~~~~~~ & 0.003 \\
  & $\eta \rho^0$ &    ~~$0.39\pm0.05$ & $0.59\pm0.08$ & $-0.10\pm0.29$ & $0.10\pm0.30$ & 1.0 \\
  & $\eta\,' \rho^0$ &  $-0.55\pm0.07$ & $-0.51\pm0.07$ & $0.12\pm0.22$ & $0.16\pm0.22$ & $-0.1$ \\
\hline
$D^+$
  & $\pi^+ \rho^0$ & 0 & $1.44\pm0.11$ & $0.78\pm1.30$ & $2.20\pm1.38$ & 0.5\\
  & $\pi^0 \rho^+$ & 0 & $-0.40\pm0.03$ & $0.90\pm0.37$ & $0.49\pm0.37$ & 0.2 \\
  & $\pi^+ \omega$ &  0 & $-0.13\pm0.03$ & $0.84\pm2.05$ & $0.74\pm2.03$ & $-0.05$ \\
  & $\pi^+ \phi$ & 0 & 0~~~~~~ & 0~~~~~~ & 0~~~~~~ &  $-0.0001$  \\
  & $\eta \rho^+$ &   ~~$1.55\pm0.26$ & $2.12\pm0.36$ & $1.22\pm0.65$ & $1.78\pm0.69$ & $-0.6$ \\
  & $\eta\,' \rho^+$ &  $-0.25\pm0.05$ & $-0.24\pm0.04$ & $0.10\pm0.12$ & $0.08\pm0.11$ & 0.5 \\
  & $K^+ \ol{K}^{*0}$ &  $-0.14\pm0.02$ & $-0.27\pm0.02$ & $-0.94\pm0.30$ & $-1.06\pm0.30$ & 0.2\\
  & $\ol{K}^0 K^{*+}$ &  $-0.06\pm0.01$ & $0.06\pm0.01$ & $-0.01\pm0.04$ & $0.10\pm0.04$ & 0.04 \\
\hline
$D_s^+$
  & $\pi^+ K^{*0}$ &  ~~$0.14\pm0.02$ & $0.24\pm0.02$ & $0.94\pm0.30$ & $1.05\pm0.30$ & $-0.1$  \\
  & $\pi^0 K^{*+}$ &  ~~$0.10\pm0.03$ & $0.04\pm0.04$ & $1.21\pm0.39$ & $1.15\pm0.40$ & $-0.2$  \\
  & $K^+ \rho^0$ &  ~~$0.10\pm0.02$ & $-0.02\pm0.02$ & $0.03\pm0.07$ & $-0.08\pm0.07$ & 0.3 \\
  & $K^0 \rho^+$ &  ~~$0.06\pm0.01$ & $-0.03\pm0.01$ & $0.01\pm0.04$ & $-0.08\pm0.04$ & 0.3 \\
  & $\eta K^{*+}$ &  $-1.03\pm0.17$ & $-0.33\pm0.06$ & $-0.61\pm0.47$ & $0.10\pm0.48$ & 1.1 \\
  & $\eta\,' K^{*+}$ &  ~~$0.25\pm0.04$ & $0.24\pm0.03$ & $-0.11\pm0.14$ & $-0.12\pm0.13$ & $-0.5$ \\
  & $K^+ \omega$ &  $-0.09\pm0.02$ & $-0.03\pm0.02$ & $-0.05\pm0.07$ & $0.01\pm0.08$ & $-2.3$ \\
  & $K^+ \phi$ & 0 & 0~~~~~~ & 0~~~~~~ & 0~~~~~~ & $-0.8$ \\
\end{tabular}
\end{ruledtabular}
\end{table}
%

\subsection{Direct \CP violation}
It has been noticed that weak penguin annihilation will receive sizable long-distance contributions from final-state rescattering. We shall assume that the long-distance $\PE_V$ and $\PE_P$ are of the same order of magnitude as $E_V$ and $E_P$ in Solution (S6), respectively. For concreteness, we take (in units of $10^{-6}$)
\be \label{eq:PEVP}
(P\!E_V)^{\rm LD} \approx (0.26\pm0.05)\,e^{i (224\pm30)^\circ}, \qquad
(P\!E_P)^{\rm LD} \approx (1.66\pm0.33)\,e^{-i (108\pm30)^\circ}.
\en
The calculated  results are shown in Table~\ref{tab:CPVP}. In comparison, the predictions given in~\cite{Qin} in general are substantially smaller than ours in magnitude.
We find several golden modes for the search of \CP violation in the $V\!P$ sector:
\be
D^0\to \pi^+\rho^-, K^+K^{*-},\qquad D^+\to K^+\overline{K}^{*0}, \eta\rho^+, \qquad
D_s^+\to \pi^+ K^{*0}, \pi^0K^{*+}.
\en
These modes are ``golden'' in the sense that they have large branching fractions and sizeable \CP asymmetries of order $1\times 10^{-3}$. It is interesting to notice that the \CP asymmetry difference defined by
\be
\Delta A_{CP}^{V\!P}\equiv a_{CP}(K^+K^{*-})-a_{CP}(\pi^+\rho^-),
\en
in analogy to $\Delta A_{CP}^{P\!P}$ defined in Eq.~(\ref{eq:LHCb2012}), is predicted to be $(-1.52\pm0.43)\times 10^{-3}$, which is very similar to the recently observed \CP violation in $D^0\to K^+K^-$ and $D^0\to\pi^+\pi^-$. It is thus desirable to first search for \CP violation in the aforementioned golden modes.

\section{Discussions and Conclusions \label{sec:summary}}

In this analysis, we have revisited two-body hadronic charmed meson decays to $P\!P$ and $V\!P$ final states, where $P$ and $V$ denote light pseudoscalar and vector mesons, respectively.  Taking flavor SU(3) symmetry as our working assumption for the Cabibbo-favored decays, we extract tree-type flavor amplitudes through a global fit to the latest experimental data.  We then discuss whether and how SU(3) symmetry breaking factors should be taken into account when moving on to the singly Cabibbo-suppressed decay modes.  We have made predictions for the branching fractions as well as the \CP asymmetries for these decay modes where we observe that the importance of penguin-type amplitudes, if present, often significantly modify the latter.

In the $PP$ sector, several SU(3) breaking effects are crucial in explaining the measured branching fractions of singly Cabibbo-suppressed decay modes, as already noticed in Ref.~\cite{Cheng:2012b}.  The $T$ and $C$ amplitudes should be scaled by a factor given under the factorization assumption.  We acknowledge that the $E$ amplitude is governed mainly by long-distance rescattering effects and, therefore, the associated symmetry breaking factors need to be obtained via a fit to the eight $D^0$ decays.  In particular, one has to distinguish between two types of $W$-exchange amplitudes: $E_d$ and $E_s$, depending upon whether it is $d \overline{d}$ or $s \overline{s}$ pair coming out of the exchange diagram.  While $|E_d|$ is about $10\%$ larger in magnitude than $|E|$ of the Cabibbo-favored modes, $|E_s|$ has two possibilities: either larger or smaller than $|E|$ by about $40\%$, as given in Eq.~\eqref{eq:EdEs}.  The above-mentioned SU(3) symmetry breaking effects are most notably successful in explaining the large disparity between ${\cal B}(D^0 \to \pi^+ \pi^-)$ and ${\cal B}(D^0 \to K^+ K^-)$.

To test among different theory models, we have proposed to have a better precision in the measurement of $a_{CP}(D^0 \to K_S K_S)$, which is primarily due to interference between $E_d$ and $E_s$ amplitudes.  We also revisit the \CP asymmetry difference between $D^0\to K^+K^-$ and $D^0\to \pi^+\pi^-$ and find two results:
$\Delta a_{CP}^{\rm dir} = (-1.14\pm0.26)\times 10^{-3}$ for Solution I and $(-1.25\pm0.25)\times 10^{-3}$ for Solution II.  Both of them are consistent with the latest LHCb result~\cite{LHCb:2019} within $1\sigma$.  We have also observed that these predictions are sensitive to the assumed contribution from weak penguin annihilation diagrams.  Comparisons with a few other works are made to highlight the distinctive features of our approach.

In contrast to the $P\!P$ sector, a global fit to the Cabibbo-favored modes in the $V\!P$ sector gives many solutions with similarly small local minima in $\chi^2$ (six of them, as listed in Table~\ref{tab:CFVPB}, when we restrict ourselves to $\chi^2_{\rm min} < 10$), revealing significant degeneracy in the current data.  These solutions can explain the Cabibbo-favored decay branching fractions well except for the $D_s^+\to \overline{K}^0K^{*+}$ and $\rho^+\eta'$ modes.  For the former, we urge the experimental colleagues to update the figure.  For the latter, an amplitude sum rule confines its branching fraction in the range $(1.6,3.9)\%$ at the $1\sigma$ level.

The above-mentioned solution degeneracy is lifted once we use them to predict for the singly Cabibbo-suppressed modes.  In the end, we find that only Solution (S3) and (S6) which have a common feature that $C_V$ and $C_P$ are close in phase in order to simultaneously explain the small ${\cal B}(D^0 \to \pi^0\omega)$ and large ${\cal B}(D^0 \to \pi^0\rho^0)$.  Another common feature is that $A_V$ and $A_P$ are comparable in size and similar in phase, in order to simultaneously explain the small ${\cal B}(D^+ \to \pi^+\omega)$ and large ${\cal B}(D^+ \to \pi^+\rho^0)$.  We note that the recent BESIII result of ${\cal B}(D_s^+ \to K^+\omega)$ is a factor of two to three smaller than our prediction, and remains an issue to be resolved.

Unlike the $P\!P$ sector, the singly Cabibbo-suppressed decay data in the $V\!P$ sector do not call for an introduction of SU(3) breaking for the $T_{V,P}$ and $C_{V,P}$ amplitudes dictated by the factorization assumption.  Instead, SU(3) breaking in $E_{V,P}$ is still required and, analogous to the $PP$ sector, one should consider different long-distance effects on diagrams with $d\overline{d}$ and $s\overline{s}$ emerging from the $W$ exchange.  A fit to eight singly Cabibbo-suppressed $D^0$ decays shows that the symmetry breaking effects are often large.  We have identified the set with the smallest SU(3) breaking in the $E$-type amplitudes ($< 50\%$) of Solution (S6) as our best solution and make predictions for the branching fractions and \CP asymmetries of the singly Cabibbo-suppressed decays.  In particular, we point out that the $D^0\to \pi^+\rho^-, K^+K^{*-}$, $D^+\to K^+\overline{K}^{*0}, \eta\rho^+$, and $D_s^+\to \pi^+ K^{*0}, \pi^0K^{*+}$ modes have sufficiently large branching fractions and \CP asymmetries at per mille level.  Interestingly, $\Delta a_{CP}^{V\!P}\equiv a_{CP}(K^+K^{*-})-a_{CP}(\pi^+\rho^-) \simeq (-1.52\pm0.43)\times 10^{-3}$, very similar to the recently observed \CP violation in $D^0\to K^+K^-$ and $D^0\to\pi^+\pi^-$.

\section*{Acknowledgments}

This research was supported in part by the Ministry of Science and Technology of R.O.C. under Grant Nos.~MOST-107-2119-M-001-034, MOST-104-2628-M-002-014-MY4 and MOST-108-2112-M-002-005-MY3.  C.W.C. is grateful to the Mainz Institute for Theoretical Physics (MITP) of the DFG Cluster of Excellence PRISMA$^+$ (Project ID 39083149) for its hospitality and its partial support during the completion of this work.


\end{document}